\documentclass[11pt,letterpaper]{article}

\usepackage{tikz}
\usepackage{tkz-graph}
\usetikzlibrary{decorations.pathreplacing}
\usepackage{epsfig}
\usepackage{graphicx}
\usepackage{amsmath}
\usepackage{amssymb}
\usepackage{algorithm}
\usepackage{algorithmic}
\usepackage{amsthm}
\usepackage[nocompress]{cite}
\usepackage{color}

\usepackage[top=1in,bottom=1in,left=1.06in,right=1.06in]{geometry}

\date{\vspace{-0.3in}}
\newtheorem{theorem}{Theorem}
\newtheorem{lemma}{Lemma}
\newtheorem{definition}{Definition}
\newcommand{\floor}[1]{\lfloor #1 \rfloor}
\renewcommand{\O}{\mathcal{O}}
\newcommand{\Qa}{\widehat{Q}}
\newcommand{\Qn}{\widetilde{Q}}
\newcommand{\si}[1]{^{(#1)}}
\newcommand{\E}{\mathbb{E}}
\renewcommand{\P}{\mathbb{P}}
\renewcommand{\exp}[1]{e^{#1}}
\newcommand{\T}{\mathcal{T}}
\newcommand{\G}{\mathcal{G}}
\newcommand{\I}{\mathbb{I}}
\renewcommand{\H}{\mathbb{H}}
\newcommand{\BREAK}{\STATE \textbf{break}}

\title{\vspace{-0.55in}
\textbf{Reconstructing a Bounded-Degree Directed Tree\\
Using Path Queries}}

\author{
Zhaosen Wang \\
CS, Purdue \\
West Lafayette, IN 47907, USA \\
\texttt{wang2055@purdue.edu}
\and
Jean Honorio \\
CS, Purdue \\
West Lafayette, IN 47907, USA \\
\texttt{jhonorio@purdue.edu}
}

\begin{document}

\maketitle

\begin{abstract}

We present a randomized algorithm for reconstructing directed rooted trees of $n$ nodes and node degree at most $d$, by asking at most $\O(dn\log^2 n)$ \emph{path queries}.
Each path query takes as input an origin node and a target node, and answers whether there is a directed path from the origin to the target.
Regarding lower bounds, we show that any randomized algorithm requires at least $\Omega(n\log n)$ queries, while any deterministic algorithm requires at least $\Omega(dn)$ queries.
Additionally, we present a $\O(dn\log^3 n)$ randomized algorithm for \emph{noisy queries}, and a $\O(dn\log^2 n)$ randomized algorithm for \emph{additive queries} on weighted trees.

\end{abstract}

\section{Introduction}

Scientists in diverse areas, such as statistics, epidemiology and economics, aim to unveil relationships within variables from collected data.
This process can be seen as the task of reconstructing a graph (i.e., finding the hidden edges of a graph) by asking queries to an oracle.
In this graph, each vertex is a variable, and each edge denotes the relationship between two variables.
For instance, in cancer research, biologists try to discover the causal relationships between genes.
By providing a specific treatment to a particular gene (origin), biologists can observe whether there is an effect in another gene (target).
This effect can be either direct (if the two genes are connected with a directed edge) or indirect (if there is a directed path from the origin to the target gene.)
In the above example, we can think of nature as the oracle that is answering whether there is a directed path from an origin to a target node.
Arguably, the cost of asking queries is very high in several application domains.
Thus, we are interested on the reconstruction of graphs that do not require the trivial $n^2$ queries for $n$ nodes (i.e., one query for every possible pair of nodes.)

Prior work on reconstructing graphs has exclusively focused on \emph{undirected} graphs.
In~\cite{Hein89}, a $\O(dn\log n)$ algorithm was provided for recovering undirected trees of $n$ nodes and maximum node degree $d$, by using queries that return the path length between two given nodes.
Authors of~\cite{Beerliova06} provide a $\O(dn)$ algorithm for reconstructing undirected trees of $n$ nodes and maximum node degree $d$, and for a query that returns the distance from a given node to every other node.
The results in~\cite{Alon04,Alon04b,Angluin04} pertain to queries that answer whether there exists at least one edge between a given set of nodes.
While~\cite{Alon04,Alon04b} focused on matchings and stars, the work of~\cite{Angluin04} provides a $\O(m\log n)$ algorithm for undirected graphs of $n$ nodes and $m$ edges.
The results in~\cite{Grebinski00,Reyzin07} pertain to queries that return the number of edges between a given set of nodes.
A $\O(dn)$ algorithm was provided in~\cite{Grebinski00} for undirected graphs of $n$ nodes and maximum node degree $d$, while a $\O(m\log n)$ algorithm was given in~\cite{Reyzin07} for undirected graphs of $n$ nodes and $m$ edges.
Other works have focused on the recovery of \emph{weighted} undirected graphs.
The work of~\cite{Culberson89} provides a $O(dn\log n)$ algorithm for recovering \emph{weighted} undirected trees of $n$ nodes and maximum node degree $d$, by using queries that return the sum of edge weights on the path between two given nodes.
The work of~\cite{Choi13,Mazzawi10} pertains to the reconstruction of \emph{weighted} undirected graphs of $n$ nodes and $m$ edges.
A $\O(m\log n)$ algorithm was provided for a query that gives the sum of edge weights between a given set of nodes.

The closest work to ours is~\cite{Jagadish13}, which provides a $\O(dn\log^2 n)$ randomized algorithm for \emph{undirected} trees of $n$ nodes and maximum node degree $d$, and for \emph{separator queries}.
A separator query takes three nodes $i$, $k$, $j$ as input, and answers whether $k$ is on the \emph{undirected} path between $i$ and $j$.
In contrast, our work pertains to \emph{directed rooted} trees.
Furthermore, we use a different type of query which we call \emph{path query}.
A path query takes an ordered pair of nodes $i$, $j$ as input, and answers whether there exists a directed path from $i$ to $j$.

We provide a randomized algorithm for reconstructing directed rooted trees of $n$ nodes and node degree at most $d$, in $\O(dn\log^2 n)$ time.
To the best of our knowledge, there is no simple reduction to transform our problem to the problem of~\cite{Jagadish13} or any of the above mentioned literature.
Our algorithm relies on the \emph{divide and conquer} approach, the use of \emph{even separators}~\cite{Chung90} and sorting.
Regarding lower bounds, we show that any randomized algorithm requires at least $\Omega(n\log n)$ queries, while any deterministic algorithm requires at least $\Omega(dn)$ queries.
We also present a $\O(dn\log^3 n)$ randomized algorithm for a \emph{noisy} regime, in which the bit that represents the oracle's answer gets flipped with some probability, by an adversary, before it is revealed to the algorithm.
Furthermore, we present a $\O(dn\log^2 n)$ randomized algorithm for reconstructing weighted trees by using \emph{additive} queries that return the sum of the edge weights on the directed
path between two given nodes.
We finish the paper by showing some negative results that provide some motivation for our assumptions.
We show that any deterministic or randomized algorithm requires at least $\Omega(n^2)$ queries in order to recover more general directed acyclic graphs.
We also show that any deterministic algorithm requires at least $\Omega(n^2)$ queries for recovering a family of sparse disconnected graphs, as well as a family of sparse connected graphs.

\section{Preliminaries} \label{sec:prelims}

In this section, we provide several formal definitions which will be useful later for the detailed description of our algorithm.
For clarity, we also provide some preliminary introduction to the main aspects of our algorithm.

Let $G=(V,E)$ be a directed acyclic graph with vertex set $V$ and edge set $E$.
For clarity, when $G$ is a directed rooted tree, we will use $T$ instead of $G$.
In this paper, we assume that $T$ has $n$ nodes, i.e., $|V|=n$.
Furthermore, we also assume that the node degree is at most $d$.
(In a directed acyclic graph, the node degree is the sum of the indegree and the outdegree of the node.)

Recall that a path in $G$ from node $i$ to node $j$ (both in $V$) is a sequence of nodes $i, x_1, x_2, \dots x_k, j$ such that $\{(i, x_1), (x_1, x_2), \dots (x_{k-1}, x_k), (x_k, j)\}$ is a subset of the edge set $E$.

Our algorithm reconstructs a directed rooted tree, by using \emph{path queries}.
Next, we formally define path queries.

\begin{definition} \label{def:query}
Let $G=(V,E)$ be a directed acyclic graph.
A \emph{path query} is a function $Q_G : V \times V \to \{0,1\}$ such that $Q_G(i, j) = 1$ if there exists a path in $G$ from $i$ to $j$, and $Q_G(i, j) = 0$ otherwise.
\end{definition}

Note that the above query only reveals a single bit of information, and it does not provide any information regarding the length of the path, thus making graph reconstruction a nontrivial task.

In this paper, we assume that the node set $V$ is known, while edge set $E$ is unknown.
Our main problem is indeed to reconstruct $E$ by using path queries.
We will use $Q(i,j)$ to denote $Q_T(i,j)$ since for our problem, the directed rooted tree $T$ is fixed (but unknown).

\begin{figure}
\begin{center}
\begin{tikzpicture}[scale=1.0]
\tikzstyle{vertex}=[circle, fill=white, draw, inner sep=0pt, minimum size=15pt]
\tikzstyle{mdpvertex}=[circle, fill=red!20, draw, inner sep=0pt, minimum size=15pt]
\node[mdpvertex][label=center:$1$](x0) at (1,1) {};
\node[mdpvertex][label=center:$2$](x1) at (2,1) {};
\node[mdpvertex][label=center:$3$](x2) at (3,1) {};
\node[mdpvertex][label=center:$4$](x3) at (4,1) {};
\node[mdpvertex][label=center:$5$](x4) at (5,1) {};
\node[vertex][label=center:$6$](x5) at (0.675,0) {};
\node[vertex][label=center:$7$](x6) at (1.325,0) {};
\node[vertex][label=center:$8$](x7) at (2,0) {};
\node[vertex][label=center:$9$](x8) at (3,0) {};
\node[vertex][label=center:$10$](x9) at (3,-1) {};
\node[vertex][label=center:$11$](x10) at (5,0) {};
\tikzset{EdgeStyle/.style={->}}
\Edge(x0)(x5)
\Edge(x0)(x6)
\Edge(x1)(x7)
\Edge(x8)(x2)
\Edge(x8)(x9)
\Edge(x4)(x10)
\tikzset{EdgeStyle/.style={->, red}}
\Edge(x1)(x0)
\Edge(x2)(x1)
\Edge(x2)(x3)
\Edge(x3)(x4)
\end{tikzpicture}
\hspace{0.4in}
\begin{tikzpicture}[scale=1.0]
\tikzstyle{vertex}=[circle, fill=white, draw, inner sep=0pt, minimum size=15pt]
\tikzstyle{mdpvertex}=[circle, fill=red!20, draw, inner sep=0pt, minimum size=15pt]
\node[mdpvertex][label=center:$1$](x0) at (1,1) {};
\node[mdpvertex][label=center:$2$](x1) at (2,1) {};
\node[mdpvertex][label=center:$3$](x2) at (3,1) {};
\node[mdpvertex][label=center:$4$](x3) at (4,1) {};
\node[mdpvertex][label=center:$5$](x4) at (5,1) {};
\node[vertex][label=center:$6$](x5) at (0.675,0) {};
\node[vertex][label=center:$7$](x6) at (1.325,0) {};
\node[vertex][label=center:$8$](x7) at (2,0) {};
\node[vertex][label=center:$9$](x8) at (3,0) {};
\node[vertex][label=center:$10$](x9) at (3,-1) {};
\node[vertex][label=center:$11$](x10) at (5,0) {};
\Edge(x0)(x5)
\Edge(x0)(x6)
\Edge(x1)(x7)
\Edge(x8)(x2)
\Edge(x8)(x9)
\Edge(x4)(x10)
\tikzset{EdgeStyle/.style={red}}
\Edge(x1)(x0)
\Edge(x2)(x1)
\Edge(x2)(x3)
\Edge(x3)(x4)
\end{tikzpicture} \\
\vspace{-0.2in}\makebox[1.7in]{}\makebox[0.4in]{(a)}\hspace{0.4in}\makebox[1.7in]{}\makebox[0.4in]{(b)}
\end{center}
\vspace{-0.25in}
\caption{(a) A directed rooted tree with \emph{multidirectional path} of nodes $1, 2, 3, 4, 5$.
Nodes and edges in the multidirectional path are shown in red.
(b) Skeleton graph of the directed rooted tree on the left, with path of nodes $1, 2, 3, 4, 5$.
Nodes and edges in the path are shown in red.}
\label{fig:multidirpath}
\end{figure}
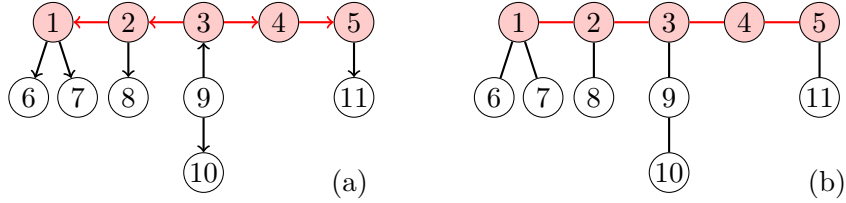

A key step in our algorithm is the recovery of what we call \emph{multidirectional paths}.
A multidirectional path consists of the \emph{directed} edges associated to an \emph{undirected} path in the skeleton graph (i.e., the graph obtained by replacing each directed edge with an undirected edge in the original graph.)
Figure~\ref{fig:multidirpath} provides a visual illustration for intuitive understanding.
Next, we formally define multidirectional paths.

\begin{definition}
Let $G=(V,E)$ be a directed acyclic graph.
A \emph{multidirectional path} of $G$ between $i$ and $j$ is a sequence of nodes $i, x_1, x_2, \dots , x_{k-1}, x_k, j$ such that each node in $V$ appears at most once in the sequence, and that there is an edge on either direction between each pair of adjacent nodes in the sequence.
That is, either $(i, x_1) \in E$ or $(x_1, i) \in E$, either $(x_1, x_2) \in E$ or $(x_2, x_1) \in E$, $\dots$ either $(x_{k-1}, x_k) \in E$ or $(x_k, x_{k-1}) \in E$, and either $(x_k, j) \in E$ or $(j, x_k) \in E$.
\end{definition}

Next, we show that for directed rooted trees, a multidirectional path between any two arbitrary nodes always exists and is unique.
More importantly, we show that a \emph{multidirectional} path is either a \emph{directed} path, or two \emph{directed} paths that share the same origin (i.e., the \emph{lowest common ancestor}.)
Later, we leverage this property for recovering multidirectional paths.

\begin{lemma} \label{lem:multidirpathfortrees}
Let $T=(V,E)$ be a directed rooted tree.
Given any two arbitrary nodes $i$ and $j$, a multidirectional path of $T$ between $i$ and $j$ always exists and is unique.
Furthermore, a multidirectional path of $T$ between $i$ and $j$, is either a path from $i$ to $j$, or a path from $j$ to $i$, or two paths (one from $k$ to $i$, and one from $k$ to $j$, for some $k \in V - \{i,j\}$.)
In this case, node $k$ is the \emph{lowest common ancestor} of $i$ and $j$.
\end{lemma}
(See Appendix~\ref{sec:detailedproofs} for detailed proofs.)

\vspace{0.1in}

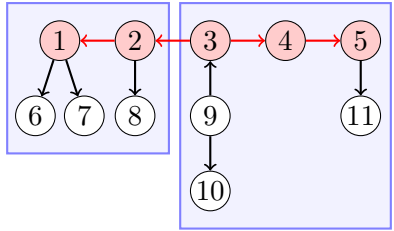
\begin{figure}
\begin{center}
\begin{tikzpicture}[scale=1.0]
\tikzstyle{vertex}=[circle, fill=white, draw, inner sep=0pt, minimum size=15pt]
\tikzstyle{mdpvertex}=[circle, fill=red!20, draw, inner sep=0pt, minimum size=15pt]
\tikzstyle{bag}=[color=blue!50, fill=blue!5, thick]
\filldraw[bag] (0.3,-0.5)rectangle (2.45,1.5);
\filldraw[bag] (2.6,-1.5)rectangle (5.4,1.5);
\node[mdpvertex][label=center:$1$](x0) at (1,1) {};
\node[mdpvertex][label=center:$2$](x1) at (2,1) {};
\node[mdpvertex][label=center:$3$](x2) at (3,1) {};
\node[mdpvertex][label=center:$4$](x3) at (4,1) {};
\node[mdpvertex][label=center:$5$](x4) at (5,1) {};
\node[vertex][label=center:$6$](x5) at (0.675,0) {};
\node[vertex][label=center:$7$](x6) at (1.325,0) {};
\node[vertex][label=center:$8$](x7) at (2,0) {};
\node[vertex][label=center:$9$](x8) at (3,0) {};
\node[vertex][label=center:$10$](x9) at (3,-1) {};
\node[vertex][label=center:$11$](x10) at (5,0) {};
\tikzset{EdgeStyle/.style={->}}
\Edge(x0)(x5)
\Edge(x0)(x6)
\Edge(x1)(x7)
\Edge(x8)(x2)
\Edge(x8)(x9)
\Edge(x4)(x10)
\tikzset{EdgeStyle/.style={->, red}}
\Edge(x1)(x0)
\Edge(x2)(x1)
\Edge(x2)(x3)
\Edge(x3)(x4)
\end{tikzpicture}
\end{center}
\vspace{-0.25in}
\caption{A directed rooted tree with multidirectional path of nodes $1, 2, 3, 4, 5$.
Nodes and edges in the multidirectional path are shown in red.
The edge $(2,3)$ in the multidirectional path is an \emph{even separator}.
Our algorithm will recursively recover each of the two generated subtrees (shown as blue boxes.)}
\label{fig:evenseparator}
\end{figure}

Our algorithm relies on the divide and conquer approach.
In order to apply the above approach in our problem, it is important to introduce the concepts of \emph{even separators} and \emph{bags}.
Next, we introduce even separators.

\begin{definition} \label{def:evenseparator}
Let $T=(V,E)$ be a directed rooted tree of bounded degree $d$ and let $n=|V|$.
An \emph{even separator} of $T$ is an edge $e \in E$ that when removed from $T$, divides $T$ into two subtrees $T_1$ and $T_2$, where each of the subtrees have a number of nodes between $n/d$ and $(d-1)n/d$.
\end{definition}

The existence of even separators is pivotal for using divide and conquer in our problem.
Corollary 2.3 in~\cite{Chung90} shows that if a graph is a bounded-degree directed tree, then an even separator exists.
For our graph reconstruction problem, once the even separator is identified, we cut the tree through the even separator.
This operation splits the tree into two subtrees.
We then recursively call the algorithm for both subtrees.
We illustrate this in Figure~\ref{fig:evenseparator}.

While even separators exist~\cite{Chung90}, it remains to know whether they can be efficiently found.
We show later that (on average) there is an even separator \emph{in the multidirectional path} between two nodes chosen independently and uniformly at random (See Theorem~\ref{thm:complexity}.)

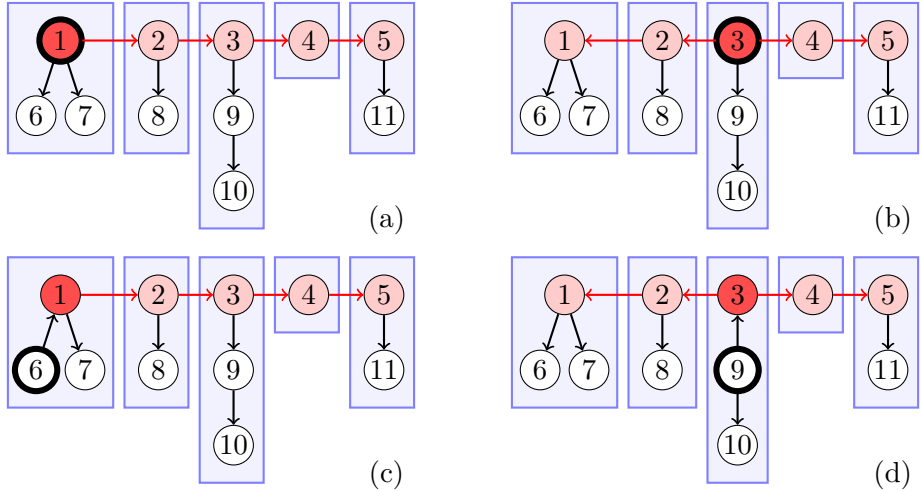
\begin{figure}
\begin{center}
\begin{tikzpicture}[scale=1.0]
\tikzstyle{vertex}=[circle, fill=white, draw, inner sep=0pt, minimum size=15pt]
\tikzstyle{mdpvertex}=[circle, fill=red!20, draw, inner sep=0pt, minimum size=15pt]
\tikzstyle{lcavertex}=[circle, fill=red!70, draw, inner sep=0pt, minimum size=15pt]
\tikzstyle{bag}=[color=blue!50, fill=blue!5, thick]
\filldraw[bag] (0,-0.5)rectangle (1.4,1.5);
\filldraw[bag] (1.55,-0.5)rectangle (2.4,1.5);
\filldraw[bag] (2.55,-1.5)rectangle (3.4,1.5);
\filldraw[bag] (3.55,0.5)rectangle (4.4,1.5);
\filldraw[bag] (4.55,-0.5)rectangle (5.4,1.5);
\node[lcavertex][minimum size=16 pt, double=black, double distance=1.5pt, label=center:$1$](x0) at (0.7,1) {};
\node[mdpvertex][label=center:$2$](x1) at (2,1) {};
\node[mdpvertex][label=center:$3$](x2) at (3,1) {};
\node[mdpvertex][label=center:$4$](x3) at (4,1) {};
\node[mdpvertex][label=center:$5$](x4) at (5,1) {};
\node[vertex][label=center:$6$](x5) at (0.375,0) {};
\node[vertex][label=center:$7$](x6) at (1.025,0) {};
\node[vertex][label=center:$8$](x7) at (2,0) {};
\node[vertex][label=center:$9$](x8) at (3,0) {};
\node[vertex][label=center:$10$](x9) at (3,-1) {};
\node[vertex][label=center:$11$](x10) at (5,0) {};
\tikzset{EdgeStyle/.style={->}}
\Edge(x0)(x5)
\Edge(x0)(x6)
\Edge(x1)(x7)
\Edge(x2)(x8)
\Edge(x8)(x9)
\Edge(x4)(x10)
\tikzset{EdgeStyle/.style={->, red}}
\Edge(x0)(x1)
\Edge(x1)(x2)
\Edge(x2)(x3)
\Edge(x3)(x4)
\end{tikzpicture}
\hspace{0.4in}
\begin{tikzpicture}[scale=1.0]
\tikzstyle{vertex}=[circle, fill=white, draw, inner sep=0pt, minimum size=15pt]
\tikzstyle{mdpvertex}=[circle, fill=red!20, draw, inner sep=0pt, minimum size=15pt]
\tikzstyle{lcavertex}=[circle, fill=red!70, draw, inner sep=0pt, minimum size=15pt]
\tikzstyle{bag}=[color=blue!50, fill=blue!5, thick]
\filldraw[bag] (0,-0.5)rectangle (1.4,1.5);
\filldraw[bag] (1.55,-0.5)rectangle (2.45,1.5);
\filldraw[bag] (2.6,-1.5)rectangle (3.4,1.5);
\filldraw[bag] (3.55,0.5)rectangle (4.4,1.5);
\filldraw[bag] (4.55,-0.5)rectangle (5.4,1.5);
\node[mdpvertex][label=center:$1$](x0) at (0.7,1) {};
\node[mdpvertex][label=center:$2$](x1) at (2,1) {};
\node[lcavertex][minimum size=16 pt, double=black, double distance=1.5pt, label=center:$3$](x2) at (3,1) {};
\node[mdpvertex][label=center:$4$](x3) at (4,1) {};
\node[mdpvertex][label=center:$5$](x4) at (5,1) {};
\node[vertex][label=center:$6$](x5) at (0.375,0) {};
\node[vertex][label=center:$7$](x6) at (1.025,0) {};
\node[vertex][label=center:$8$](x7) at (2,0) {};
\node[vertex][label=center:$9$](x8) at (3,0) {};
\node[vertex][label=center:$10$](x9) at (3,-1) {};
\node[vertex][label=center:$11$](x10) at (5,0) {};
\tikzset{EdgeStyle/.style={->}}
\Edge(x0)(x5)
\Edge(x0)(x6)
\Edge(x1)(x7)
\Edge(x2)(x8)
\Edge(x8)(x9)
\Edge(x4)(x10)
\tikzset{EdgeStyle/.style={->, red}}
\Edge(x1)(x0)
\Edge(x2)(x1)
\Edge(x2)(x3)
\Edge(x3)(x4)
\end{tikzpicture} \\
\vspace{-0.2in}\makebox[1.85in]{}\makebox[0.4in]{(a)}\hspace{0.4in}\makebox[1.85in]{}\makebox[0.4in]{(b)}\vspace{0.1in} \\
\begin{tikzpicture}[scale=1.0]
\tikzstyle{vertex}=[circle, fill=white, draw, inner sep=0pt, minimum size=15pt]
\tikzstyle{mdpvertex}=[circle, fill=red!20, draw, inner sep=0pt, minimum size=15pt]
\tikzstyle{lcavertex}=[circle, fill=red!70, draw, inner sep=0pt, minimum size=15pt]
\tikzstyle{bag}=[color=blue!50, fill=blue!5, thick]
\filldraw[bag] (0,-0.5)rectangle (1.4,1.5);
\filldraw[bag] (1.55,-0.5)rectangle (2.4,1.5);
\filldraw[bag] (2.55,-1.5)rectangle (3.4,1.5);
\filldraw[bag] (3.55,0.5)rectangle (4.4,1.5);
\filldraw[bag] (4.55,-0.5)rectangle (5.4,1.5);
\node[lcavertex][label=center:$1$](x0) at (0.7,1) {};
\node[mdpvertex][label=center:$2$](x1) at (2,1) {};
\node[mdpvertex][label=center:$3$](x2) at (3,1) {};
\node[mdpvertex][label=center:$4$](x3) at (4,1) {};
\node[mdpvertex][label=center:$5$](x4) at (5,1) {};
\node[vertex][minimum size=16 pt, double=black, double distance=1.5pt, label=center:$6$](x5) at (0.375,0) {};
\node[vertex][label=center:$7$](x6) at (1.025,0) {};
\node[vertex][label=center:$8$](x7) at (2,0) {};
\node[vertex][label=center:$9$](x8) at (3,0) {};
\node[vertex][label=center:$10$](x9) at (3,-1) {};
\node[vertex][label=center:$11$](x10) at (5,0) {};
\tikzset{EdgeStyle/.style={->}}
\Edge(x5)(x0)
\Edge(x0)(x6)
\Edge(x1)(x7)
\Edge(x2)(x8)
\Edge(x8)(x9)
\Edge(x4)(x10)
\tikzset{EdgeStyle/.style={->, red}}
\Edge(x0)(x1)
\Edge(x1)(x2)
\Edge(x2)(x3)
\Edge(x3)(x4)
\end{tikzpicture}
\hspace{0.4in}
\begin{tikzpicture}[scale=1.0]
\tikzstyle{vertex}=[circle, fill=white, draw, inner sep=0pt, minimum size=15pt]
\tikzstyle{mdpvertex}=[circle, fill=red!20, draw, inner sep=0pt, minimum size=15pt]
\tikzstyle{lcavertex}=[circle, fill=red!70, draw, inner sep=0pt, minimum size=15pt]
\tikzstyle{bag}=[color=blue!50, fill=blue!5, thick]
\filldraw[bag] (0,-0.5)rectangle (1.4,1.5);
\filldraw[bag] (1.55,-0.5)rectangle (2.45,1.5);
\filldraw[bag] (2.6,-1.5)rectangle (3.4,1.5);
\filldraw[bag] (3.55,0.5)rectangle (4.4,1.5);
\filldraw[bag] (4.55,-0.5)rectangle (5.4,1.5);
\node[mdpvertex][label=center:$1$](x0) at (0.7,1) {};
\node[mdpvertex][label=center:$2$](x1) at (2,1) {};
\node[lcavertex][label=center:$3$](x2) at (3,1) {};
\node[mdpvertex][label=center:$4$](x3) at (4,1) {};
\node[mdpvertex][label=center:$5$](x4) at (5,1) {};
\node[vertex][label=center:$6$](x5) at (0.375,0) {};
\node[vertex][label=center:$7$](x6) at (1.025,0) {};
\node[vertex][label=center:$8$](x7) at (2,0) {};
\node[vertex][minimum size=16 pt, double=black, double distance=1.5pt, label=center:$9$](x8) at (3,0) {};
\node[vertex][label=center:$10$](x9) at (3,-1) {};
\node[vertex][label=center:$11$](x10) at (5,0) {};
\tikzset{EdgeStyle/.style={->}}
\Edge(x0)(x5)
\Edge(x0)(x6)
\Edge(x1)(x7)
\Edge(x8)(x2)
\Edge(x8)(x9)
\Edge(x4)(x10)
\tikzset{EdgeStyle/.style={->, red}}
\Edge(x1)(x0)
\Edge(x2)(x1)
\Edge(x2)(x3)
\Edge(x3)(x4)
\end{tikzpicture} \\
\vspace{-0.2in}\makebox[1.85in]{}\makebox[0.4in]{(c)}\hspace{0.4in}\makebox[1.85in]{}\makebox[0.4in]{(d)}
\end{center}
\vspace{-0.25in}
\caption{Four different directed rooted trees, and their \emph{bags} with respect to the multidirectional path of nodes $1, 2, 3, 4, 5$.
Nodes and edges in the multidirectional path are shown in red.
The lowest common ancestor of nodes $1$ and $5$ in the multidirectional path is shown in darker red.
A tree root is shown as a thick circle.
Bags are shown as blue boxes.
Note that a multidirectional path can be a path (a,c) or two paths (b,d).
The root can be in the multidirectional path (a,b).
Otherwise, the root can be an ancestor of a node in the multidirectional path (c,d).}
\label{fig:bag}
\end{figure}

In what follows, we formally define \emph{bags}, which are also important for the divide and conquer approach taken here.

\begin{definition} \label{def:bag}
Let $T=(V,E)$ be a directed rooted tree, and $S$ be the set of edges in a multidirectional path of $T$.
Define $T_S=(V, E - S)$ as the subgraph of $T$ after we remove all edges in $S$.
A \emph{bag} with respect to a node $i$ in a multidirectional path with edges $S$, is a subset of nodes in $V$ that contains $i$ and all the nodes that are reachable from $i$ in the (undirected) skeleton graph of $T_S$.
\end{definition}

Intuitively speaking, we can think of edges in a tree as ``ropes''.
If we ``nail'' all nodes of a multidirectional path into the ``wall'', then all other nodes will ``hang'' from one of the nodes in the path.
Nodes that hang from the same particular node belong to the same bag.
We include a visual example in Figure~\ref{fig:bag}.

In our algorithm, we recover \emph{exactly} all the directed edges \emph{in a multidirectional path}.
Bags are used to count the number of nodes associated to each node in the multidirectional path (without the need to recover all the directed edges.)
For each edge in the multidirectional path, one can then count the number of nodes on the two subtrees that would be generated if we were to cut the tree through the given edge.
This process is used for identifying even separators.

Finally, our algorithm also performs sorting of nodes with a properly defined order relation, which is used for instance in the recovery of the directed edges in multidirectional paths.

\begin{definition} \label{def:orderrelation}
Define the order relation of two nodes $i$ and $j$ as follows.
If $Q(i, j) = 1$ we say that $i$ is ``less than'' $j$, and ``greater than'' otherwise.
\end{definition}

\section{Algorithm} \label{sec:algorithm}

In this section, we present our randomized algorithm and analyze its time complexity.
Our algorithm is similar in spirit to~\cite{Jagadish13} which applies to \emph{undirected} trees and \emph{separator queries}.
In this paper, we focus on \emph{directed rooted} trees and \emph{path queries}.
We remind the reader that path queries only reveal a single bit of information, and they do not provide any information regarding the length of the path.
(We discuss noisy and additive extensions in Section~\ref{sec:extensions}.)

\begin{figure*}
\vspace{-0.25in}
\begin{center}
\begin{tikzpicture}[scale=1.0]
\tikzstyle{mainalgo}=[rectangle, fill=red!20, draw, inner sep=2pt, minimum height=30pt]
\tikzstyle{subroutine}=[rectangle, fill=white, draw, inner sep=2pt, minimum height=30pt]
\node[mainalgo][minimum width=135pt, text width=130pt, align=center](reconstructtree) at (5.5,4.1) {Alg.~\ref{alg:reconstructtree}: Reconstruct tree\\$\O(dn\log^2 n)$};
\node[subroutine][minimum width=215pt, text width=210pt, align=center](findmultidirpath) at (0.7,2.5) {Alg.~\ref{alg:findmultidirpath}: Reconstruct multidirectional path\\$\O(n\log n)$};
\node[subroutine][minimum width=95pt, text width=90pt, align=center](findbag) at (6.8,2.5) {Alg.~\ref{alg:findbag}: Find bag\\$\O(\log n)$};
\node[subroutine][minimum width=95pt, text width=90pt, align=center](splittree) at (10.9,2.5) {Alg.~\ref{alg:splittree}: Split tree\\$\O(n)$};
\node[subroutine][minimum width=195pt, text width=190pt, align=center](findcommonancestor) at (0.7,0.5) {Alg.~\ref{alg:findcommonancestor}: Find lowest common ancestor\\$\O(n\log n)$};
\node[subroutine][minimum width=150pt, text width=145pt, align=center](findroot) at (8.9,0.5) {Alg.~\ref{alg:findroot}: Find path from root\\$\O(n\log n)$};
\tikzset{EdgeStyle/.style={->}}
\Edge(reconstructtree)(findmultidirpath)
\Edge(reconstructtree)(findbag)
\Edge(reconstructtree)(splittree)
\Edge[label=if multidirectional path consists of two paths, labelstyle=right](findmultidirpath)(findcommonancestor)
\Edge(findcommonancestor)(findroot)
\path[->, thick] (reconstructtree) edge [loop left] node {} ();
\end{tikzpicture}
\end{center}
\vspace{-0.25in}
\caption{Main algorithm (in red), subroutines, and their time complexity.}
\label{fig:algorithm}
\end{figure*}
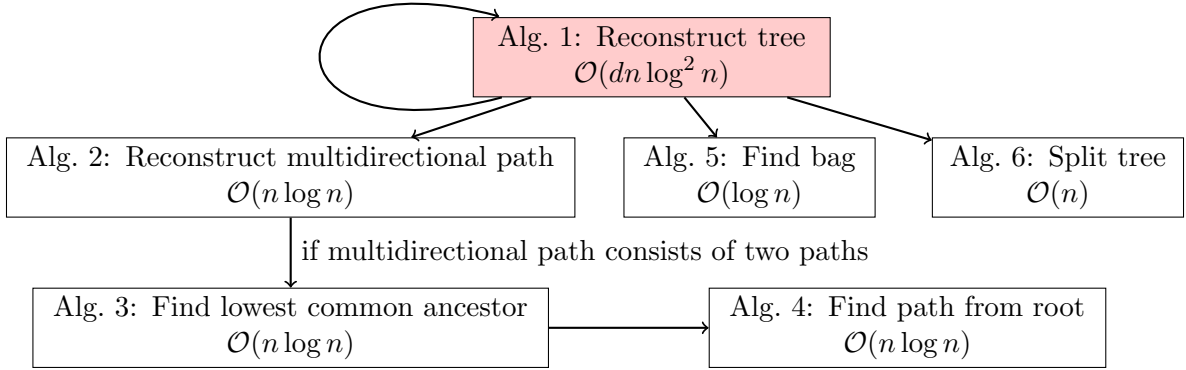

In what follows, we explain our divide and conquer approach in our main Algorithm~\ref{alg:reconstructtree}.
A high-level overview is shown in Figure~\ref{fig:algorithm}.
(See Appendix~\ref{sec:detailedalgorithms} for detailed algorithms.)
First, we randomly pick two nodes and recover the sequence of nodes in the \emph{multidirectional path} between those two randomly chosen nodes (Algorithm~\ref{alg:findmultidirpath}.)
Then, we divide the rest of the nodes (not in the multidirectional path) into \emph{bags} defined by the multidirectional path (Algorithm~\ref{alg:findbag}.)
Later, we count the number of nodes on each bag, and determine if there exists an even separator in the multidirectional path.
We repeat the whole procedure (i.e., from choosing a new pair of random nodes) until we find an even separator.
Finally, we cut the tree through the even separator, thus effectively splitting the tree into two subtrees (Algorithm~\ref{alg:splittree}.)
We then recursively call our Algorithm~\ref{alg:reconstructtree} for both subtrees (until the input tree contains a single node.)
On a more technical side, reconstructing a multidirectional path that consists of two paths that share the same origin, is more involved than reconstructing single paths.
The former requires finding the lowest common ancestor of the two randomly chosen nodes (Algorithm~\ref{alg:findcommonancestor}) as well as finding the root of the tree (Algorithm~\ref{alg:findroot}.)

In our main Algorithm~\ref{alg:reconstructtree}, finding out the bag of a particular node with respect to a \emph{directed} path is relatively easier than with respect to a \emph{multidirectional} path.
Note that by Lemma~\ref{lem:multidirpathfortrees}, a \emph{multidirectional} path is either a \emph{directed} path, or two \emph{directed} paths that share the same origin (i.e., the lowest common ancestor.)
In Algorithm~\ref{alg:reconstructtree}, we simplify the bag assignment task, by splitting a \emph{multidirected} path (that is not a single path) into its two constituent \emph{directed} paths.
The algorithm first recovers the lowest common ancestor, and then breaks the multidirectional path into two paths.
For instance, the multidirectional path $1,2,3,4,5$ in Figure~\ref{fig:bag}(d) with lowest common ancestor $3$ would become two paths $3,2,1$ and $3,4,5$.

Next, we explain each subroutine used in our main Algorithm~\ref{alg:reconstructtree}.
The first key step is to recover the multidirectional path between two randomly chosen nodes.
Algorithm~\ref{alg:findmultidirpath} recovers the sequence of nodes in the multidirectional path between any two nodes.

The idea behind Algorithm~\ref{alg:findmultidirpath} is as follows.
Let $i,j \in V$ be two arbitrary nodes in the directed rooted tree $T=(V,E)$.
Recall that some \emph{multidirectional} paths are a single \emph{directed} path.
We can easily detect this case by asking whether there is a path from $i$ to $j$ (i.e., whether $Q(i,j)=1$), or whether there is a path from $j$ to $i$ (i.e., whether $Q(j,i)=1$.)
Without loss of generalization, assume there is a path in the directed rooted tree $T$ from $i$ to $j$.
(We could similarly assume that there is a path from $j$ to $i$.)
We can find out the set of all nodes $\{x_1, x_2, \dots x_k\}$ on the path from $i$ to $j$, by using path queries.
For all nodes $k \in V - \{i,j\}$, we ask the oracle about $Q(i, k)$ and $Q(k, j)$.
Note that $k$ is on the path from $i$ to $j$, if and only if $Q(i, k)=Q(k, j)=1$.
After finding out the set of all nodes $\{x_1, x_2, \dots x_k\}$ on the path from $i$ to $j$, it remains to sort the nodes in order to obtain the correct sequence, thus recovering the path.
We sort the list of nodes $\{i, x_1, x_2, \dots x_k, j\}$ by using the order relation given in Definition~\ref{def:orderrelation}.

Some \emph{multidirectional} paths consist of two \emph{directed} paths that share the same origin $m$.
In this case, Algorithm~\ref{alg:findmultidirpath} first recovers the lowest common ancestor $m$ and then reconstructs two \emph{directed} paths: one from $m$ to $i$, and one from $m$ to $j$, by following the approach explained before.
In Algorithm~\ref{alg:findmultidirpath}, reconstructing a multidirectional path that consists of two paths requires finding the lowest common ancestor of the two randomly chosen nodes.
Our Algorithm~\ref{alg:findcommonancestor} finds the lowest common ancestor of a multidirectional path between any two arbitrary nodes.

Algorithm~\ref{alg:findcommonancestor} works as follows.
Let $i,j \in V$ be two arbitrary nodes of the directed rooted tree $T=(V,E)$.
First, we recover the \emph{directed} path from the root to $i$.
We assume that the order of the nodes in the above path follow the order relation given in Definition~\ref{def:orderrelation}.
Thus, the tree root is the first element on such path.
Then, we iterate through all nodes in the path, in order to find the last ancestor of $j$ in the path from the root.
This last ancestor is indeed the lowest common ancestor of $i$ and $j$.
In Algorithm~\ref{alg:findcommonancestor}, in order to find the lowest common ancestor, one has to find the root of the directed tree.
Our Algorithm~\ref{alg:findroot} identifies the path from the root to a given arbitrary node.

The inner workings of Algorithm~\ref{alg:findroot} are as follows.
Let $i \in V$ be an arbitrary node in the directed rooted tree $T = (V,E)$.
For each node $j \in V - \{i\}$, we ask the oracle about $Q(j,i)$.
If $Q(j,i)=1$ then there is a path from $j$ to $i$, and therefore we add node $j$ to the list of nodes that reach $i$.
In order to recover the \emph{directed} path from the root to $i$, we sort the aforementioned list of nodes, by using the order relation given in Definition~\ref{def:orderrelation}.
That is, the first element on the sorted list is the tree root.

The second key step in our main Algorithm~\ref{alg:reconstructtree} is to divide the nodes (which are not in the multidirectional path) into bags defined by the multidirectional path.
As argued before, if the \emph{multidirectional} path consists of two \emph{directed} paths, then Algorithm~\ref{alg:reconstructtree} breaks the \emph{multidirectional} path into its two constituent \emph{directed} paths, by using the least common ancestor.
Thus, the bag assigment task needs only to consider directed paths, as we do in Algorithm~\ref{alg:findbag}.

Here we give an intuitive explanation of the bag assignment task in Algorithm~\ref{alg:findbag}.
For instance, consider finding out the bag of node $10$ in the directed path $1,2,3,4,5$ in Figure~\ref{fig:bag}(a).
We see that nodes $1$, $2$ and $3$ are all ancestors of node $10$ (i.e., $Q(1,10)=Q(2,10)=Q(3,10)=1$.)
We also see that nodes $4$ and $5$ are not ancestors of node $10$ (i.e., $Q(4,10)=Q(5,10)=0$.)
Note that node $10$ belongs to the bag of node $3$.
The above suggests that the task of finding out the bag of node $10$ can be done by searching for a node $i$ for which $Q(i,10)=1$ and $Q(j,10)=0$ where $(i,j)$ is an edge in the path.
This can be efficiently done by performing binary search.

There are two exceptions to the above rule: when queries for all nodes in the path return $1$, and when queries for all nodes in the path return $0$.
As an example for the first case (when all queries return $1$), consider finding out the bag of node $11$ in the directed path $1,2,3,4,5$ in Figure~\ref{fig:bag}(a).
We see that nodes $1$, $2$, $3$, $4$ and $5$ are all ancestors of node $11$ (i.e., $Q(1,11)=Q(2,11)=Q(3,11)=Q(4,11)=Q(5,11)=1$.)
In this case, we assign node $11$ to the bag of the last node in the path, i.e., $5$.
As an example for the second case (when all queries return $0$), consider finding out the bag of node $10$ in the directed path $3,4,5$ in Figure~\ref{fig:bag}(d).
We see that neither nodes $3$, $4$ or $5$ are ancestors of node $10$ (i.e., $Q(3,10)=Q(4,10)=Q(5,10)=0$.)
In this case, we assign node $10$ to the bag of the lowest common ancestor, i.e., $3$.
Fortunately, all the cases analyzed above are naturally handled by binary search.

The final step in our main Algorithm~\ref{alg:reconstructtree} is to cut the tree through the even separator, which splits the tree into two subtrees.
Our Algorithm~\ref{alg:splittree} splits a directed rooted tree into two subtrees, by cutting the original tree through any arbitrary edge.

The idea behind Algorithm~\ref{alg:splittree} is as follows.
Let $e = (i,j) \in E$ be an arbitrary edge in the directed rooted tree $T=(V,E)$.
Let $T_1$ and $T_2$ be two subtrees that result from removing $e$ from $T$.
Note that every node must belong to either $T_1$ or $T_2$.
Without loss of generality, let $i \in T_1$ and $j \in T_2$.
Since $T$ is a directed rooted tree, $j$ has one parent in $T$, which is indeed $i$.
Removing $(i, j)$ from $T$ makes $j$ have no parents in $T_2$.
Therefore, $j$ is the root of $T_2$.
For all nodes $k \in V - \{j\}$, we ask the oracle about $Q(j, k)$.
If $Q(j, k) = 1$, then $k$ belongs to $T_2$, otherwise $k$ belongs to $T_1$.

We finish the section by analyzing the time complexity of our randomized algorithm.

\begin{theorem} \label{thm:complexity}
Algorithm~\ref{alg:reconstructtree} takes $\O(dn \log^2 n)$ expected time, in order to reconstruct a directed rooted tree of $n$ nodes and node degree at most $d$.
Furthermore, for a fixed probability of error $\delta \in (0,1)$, Algorithm~\ref{alg:reconstructtree} takes at most $\O(\frac{1}{\delta} \, dn \log^2 n)$ time, with probability at least $1-\delta$.
\end{theorem}
(See Appendix~\ref{sec:detailedproofs} for detailed proofs.)

\section{Lower Bound and Extensions} \label{sec:extensions}

In this section, we study lower bounds for reconstructing directed rooted trees from path queries.
We also extend our original algorithm for the case of noisy queries, as well as the reconstruction of weighted trees from additive queries.
Finally, we provide negative results for directed acyclic graphs that provide some motivation for our assumptions.

\subsection{Lower Bounds}

Here, we present lower bounds for reconstructing directed rooted trees from path queries.

\begin{theorem} \label{thm:lowerboundrandomized}
In order to reconstruct a directed rooted tree of $n$ nodes and node degree at most $d$, any \emph{randomized} algorithm requires at least $\Omega((1-\delta) \, n\log n)$ queries, otherwise it would fail with probability at least $\delta$.
\end{theorem}

\begin{theorem} \label{thm:lowerbounddeterministic}
In order to reconstruct a directed rooted tree of $n$ nodes and node degree at most $d$, any \emph{deterministic} algorithm requires at least $\Omega(dn)$ queries.
\end{theorem}

Given the above results and Theorem~\ref{thm:complexity}, our algorithm is only a factor of $\O(d\log n)$ from the lower bound for any \emph{randomized} algorithm, and a factor of $\O(\log^2 n)$ from the lower bound for any \emph{deterministic} algorithm.
Both of these factors are small when compared to the time complexity of our algorithm, which is $\O(dn\log^2 n)$.

\subsection{Noisy Queries}

Here, we analyze a \emph{noisy} regime, in which the bit that represents the oracle's answer gets flipped with some probability, by an adversary, before it is revealed to the algorithm.
Next, we formally define noisy queries.

\begin{definition} \label{def:noisyquery}
Let $G=(V,E)$ be a directed acyclic graph, and let $Q_G$ be a path query.
A \emph{noisy path query} with noise parameter $\varepsilon \in (0,1/2)$ is a function $\Qn_G : V \times V \to \{0,1\}$ such that $\Qn_G(i, j) = Q_G(i,j)$ with probability $1-\varepsilon$, and $\Qn_G(i, j) = 1 - Q_G(i,j)$ with probability $\varepsilon$.
\end{definition}

In order to make use of our original algorithm, we proceed with the following strategy.
Algorithm~\ref{alg:reconstructnoisytree} works as follows.
For each node pair, we will perform majority voting on $m$ noisy path queries.
If $m$ is large enough, noise will be removed with high probability.
(See Appendix~\ref{sec:detailedalgorithms} for details.)

The above opens up a question on the number of queries $m$ per node pair, that are sufficient to guarantee graph recovery success.
A second question is whether $m$ depends on the number of nodes $n$ and maximum node degree $d$, thus affecting the time complexity of our randomized algorithm.
The following theorem answers both questions.

\begin{theorem} \label{thm:complexitynoisy}
For a fixed probability of error $\delta \in (0,1)$ and noise parameter $\varepsilon \in (0,1/2)$, Algorithm~\ref{alg:reconstructnoisytree} takes at most $\O(\frac{1}{\delta} \, \frac{1}{(1/2-\varepsilon)^2} \, dn \log^2 n(\log d + \log n + \log\frac{1}{\delta}))$ time, with probability at least $1-\delta$, in order to reconstruct a directed rooted tree of $n$ nodes and node degree at most $d$, provided that the number of queries per node pair fulfills $m \in \Theta(\frac{1}{(1/2-\varepsilon)^2} \, (\log d + \log n + \log\frac{1}{\delta}))$.
\end{theorem}

In practice we do not need to know the exact value of the noise parameter $\varepsilon$.
A lower bound of $\varepsilon$ suffices to define the number of queries $m$ per node pair.

\subsection{Additive Queries on Weighted Trees}

Here, we focus on the reconstruction of \emph{weighted} directed rooted trees by using additive queries.
An additive path query returns the sum of the edge weights on the directed path between two given nodes, if such a path exists, or zero otherwise.
Next, we formally define additive path queries.

\begin{definition}
Let $T=(V,E,W)$ be a weighted directed rooted tree, with positive weights for each edge, i.e., $w_{i,j} > 0$ for all $(i,j) \in E$, and $w_{i,j} = 0$ for all $(i,j) \notin E$.
An \emph{additive path query} is a function $\Qa_T : V \times V \to [0,+\infty)$ such that if there exists a path in $T$ from $i$ to $j$ then $\Qa_T(i, j)$ is the sum of the weights of the edges in the path, and $\Qa_T(i, j) = 0$ otherwise.
\end{definition}

Note that the above query reveals much more information compared to the path query in Definition~\ref{def:query} which only reveals a single bit of information.
In Algorithm~\ref{alg:reconstructweightedtree}, our strategy is to convert the \emph{additive query} problem into our original problem for recovering the edge set.
Afterwards, we recover the edge weights by calling the \emph{additive} queries for each edge.
(See Appendix~\ref{sec:detailedalgorithms} for details.)

The time complexity of the above algorithm is as follows.

\begin{theorem} \label{thm:complexityweighted}
Algorithm~\ref{alg:reconstructweightedtree} takes $\O(dn \log^2 n)$ expected time, in order to reconstruct a weighted directed rooted tree of $n$ nodes and node degree at most $d$.
Furthermore, for a fixed probability of error $\delta \in (0,1)$, Algorithm~\ref{alg:reconstructweightedtree} takes at most $\O(\frac{1}{\delta} \, dn \log^2 n)$ time, with probability at least $1-\delta$.
\end{theorem}

\subsection{Negative Results for Directed Acyclic Graphs}

As argued before, the cost of asking queries is very high in several application domains.
Thus, we are interested on the reconstruction of graphs that do not require the trivial $n^2$ queries for $n$ nodes (i.e., one query for every possible pair of nodes.)
A natural question is whether more general directed acyclic graphs could be recovered efficiently by asking less than $\Omega(n^2)$ queries.
Here, we provide a negative answer for the above.

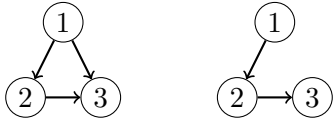
\begin{figure}
\begin{center}
\begin{tikzpicture}[scale=1.0]
\tikzstyle{vertex}=[circle, fill=white, draw, inner sep=0pt, minimum size=15pt]
\node[vertex][label=center:$1$](x0) at (1.5,2) {};
\node[vertex][label=center:$2$](x1) at (1,1) {};
\node[vertex][label=center:$3$](x2) at (2,1) {};
\tikzset{EdgeStyle/.style={->}}
\Edge(x0)(x1)
\Edge(x0)(x2)
\Edge(x1)(x2)
\end{tikzpicture}
\hspace{0.4in}
\begin{tikzpicture}[scale=1.0]
\tikzstyle{vertex}=[circle, fill=white, draw, inner sep=0pt, minimum size=15pt]
\node[vertex][label=center:$1$](x0) at (1.5,2) {};
\node[vertex][label=center:$2$](x1) at (1,1) {};
\node[vertex][label=center:$3$](x2) at (2,1) {};
\tikzset{EdgeStyle/.style={->}}
\Edge(x0)(x1)
\Edge(x1)(x2)
\end{tikzpicture}
\end{center}
\vspace{-0.25in}
\caption{Two directed acyclic graphs that produce the same answers when using path queries.}
\label{fig:nonidentifiabledag}
\end{figure}

First note that some directed acyclic graphs are non-identifiable by using path queries.
For instance, consider the two graphs shown in Figure~\ref{fig:nonidentifiabledag}.
In both graphs, we have that $Q(1,2)=Q(2,3)=Q(1,3)=1$.
Thus, by using path queries, it is impossible to discern whether the edge $(1,3)$ exists or not.
Next, we formalize the above intuition.

\begin{definition}
Let $G=(V,E)$ be a directed acyclic graph.
We say that an edge $(i,j) \in E$ is \emph{transitive} if there exists a directed path from $i$ to $j$ of length greater than 1.
\end{definition}

In Figure~\ref{fig:nonidentifiabledag}, edge $(1,3)$ is transitive, since there is a directed path $1,2,3$ (i.e., a directed path of length 2.)
Transitive edges are not possible to be recovered by using path queries for the following reason.
Let $i$ and $j$ be two fixed nodes.
Assume there is a directed path from $i$ to $j$ of length greater than 1.
Due to this path, we have that $Q(i,j)=1$, regardless of whether $(i,j) \in E$ or $(i,j) \notin E$.
A trivial algorithm for reconstructing directed acyclic graphs of $n$ nodes, without transitive edges, is to ask $n^2$ queries (i.e., one query for every possible pair of nodes.)
Next, we show that this trivial algorithm is indeed optimal.

\begin{theorem} \label{thm:lowerbounddags}
In order to reconstruct a directed acyclic graph of $n$ nodes, \emph{without transitive edges}, any \emph{deterministic} algorithm requires at least $\Omega(n^2)$ queries.
Furthermore, any \emph{randomized} algorithm requires at least $\Omega((1-\delta) \, n^2)$, otherwise it would fail with probability at least $\delta$.
\end{theorem}

Recall that our algorithm pertains to directed rooted trees with a maximum node degree.
These graphs are not only sparse but also \emph{weakly} connected (i.e., their undirected skeleton graphs are connected.)
One could ask whether connectedness is a necessary condition, and whether sparsity makes graph reconstruction easier.
Next, we show that an algorithm requires $\Omega(n^2)$ queries for recovering a family of sparse disconnected graphs, as well as a family of sparse connected graphs.

\begin{theorem} \label{thm:lowerboundedge}
In order to reconstruct a \emph{sparse disconnected} directed acyclic graph of $n$ nodes, any \emph{deterministic} algorithm requires at least $\Omega(n^2)$ queries.
\end{theorem}

\begin{theorem} \label{thm:lowerboundsparse}
In order to reconstruct a \emph{sparse connected} directed acyclic graph of $n$ nodes, any \emph{deterministic} algorithm requires at least $\Omega(n^2)$ queries.
\end{theorem}

\section{Concluding Remarks} \label{sec:conclusions}

There are several ways of extending this research.
The analysis of the reconstruction of other families of graphs in $\O(n\log n)$ time would be of great interest.
Given our results for directed rooted trees, it would be interesting to analyze other families of sparse connected graphs, such as graphs with bounded tree-width as well as graphs with bounded arboricity.

\bibliographystyle{plain}
\bibliography{references}

\begin{thebibliography}{10}

\bibitem{Akra98}
Mohamad Akra and Louay Bazzi.
\newblock On the solution of linear recurrence equations.
\newblock {\em Computational Optimization and Applications}, 10(2):195--210,
  1998.

\bibitem{Alon04b}
Noga Alon and Vera Asodi.
\newblock Learning a hidden subgraph.
\newblock {\em International Colloquium on Automata, Languages and Programming,
  Lecture Notes in Computer Science}, 3142:110--121, 2004.

\bibitem{Alon04}
Noga Alon, Richard Beigel, Simon Kasif, Steven Rudich, and Benny Sudakov.
\newblock Learning a hidden matching.
\newblock {\em SIAM Journal on Computing}, 33(2):487--–501, 2004.

\bibitem{Angluin04}
Dana Angluin and Jiang Chen.
\newblock Learning a hidden graph using $\uppercase{O}(\log n)$ queries per
  edge.
\newblock {\em Conference on Learning Theory}, pages 210--223, 2004.

\bibitem{Beerliova06}
Zuzana Beerliova, Felix Eberhard, Thomas Erlebach, Alexander Hall, Michael
  Hoffmann, Mat\'{u}\v{s} {Mihal'\'{a}k}, and {L.~Shankar} Ram.
\newblock Network discovery and verification.
\newblock {\em IEEE Journal on Selected Areas in Communications},
  24(12):2168--2181, 2006.

\bibitem{Choi13}
Sung-Soon Choi.
\newblock Polynomial time optimal query algorithms for finding graphs with
  arbitrary real weights.
\newblock {\em Conference on Learning Theory}, pages 1--22, 2013.

\bibitem{Chung90}
Fan Chung.
\newblock Separator theorems and their applications.
\newblock {\em Paths, Flows, and VLSI-Layout}, pages 17--34, 1990.

\bibitem{Cover06}
Thomas Cover and Joy Thomas.
\newblock {\em Elements of Information Theory}.
\newblock John Wiley \& Sons, 2nd edition, 2006.

\bibitem{Culberson89}
Joseph Culberson and Piotr Rudnicki.
\newblock A fast algorithm for reconstructing trees from distance matrices.
\newblock {\em Information Processing Letters}, 30(4):215--220, 1989.

\bibitem{Grebinski00}
Vladimir Grebinski and Gregory Kucherov.
\newblock Optimal reconstruction of graphs under the additive model.
\newblock {\em Algorithmica}, 28(1):104--124, 2000.

\bibitem{Hein89}
Jotun Hein.
\newblock An optimal algorithm to reconstruct trees from additive distance
  data.
\newblock {\em Bulletin of Mathematical Biology}, 51(5):597--603, 1989.

\bibitem{Jagadish13}
M.~Jagadish and Anindya Sen.
\newblock Learning a bounded-degree tree using separator queries.
\newblock {\em Algorithmic Learning Theory, Lecture Notes in Computer Science},
  8139:188--202, 2013.

\bibitem{Mazzawi10}
Hanna Mazzawi.
\newblock Optimally reconstructing weighted graphs using queries.
\newblock {\em ACM-SIAM Symposium on Discrete Algorithms}, pages 608--615,
  2010.

\bibitem{Reyzin07}
Lev Reyzin and Nikhil Srivastava.
\newblock Learning and verifying graphs using queries with a focus on edge
  counting.
\newblock {\em Algorithmic Learning Theory, Lecture Notes in Computer Science},
  4754:285--297, 2007.

\end{thebibliography}

\clearpage

\appendix

\section{Detailed Algorithms} \label{sec:detailedalgorithms}

In this section, we show the pseudocode for all the algorithms in our manuscript.
First, we present our main Algorithm~\ref{alg:reconstructtree} for reconstructing a directed rooted tree by using path queries.

\begin{algorithm}[H]
\caption{Reconstruct tree}
\label{alg:reconstructtree}
\begin{algorithmic}[1]
\STATE \textbf{Input:} vertex set $V$ of the directed rooted tree $T$
\IF{$|V|=1$}
    \STATE $E \leftarrow \emptyset$
\ELSE
  \WHILE{true} \label{lin:round}
      \STATE Pick $i, j \in V$ independently and uniformly at random
      \STATE (multidirpath, lowestancestoridx) $\leftarrow$ \textbf{Reconstruct multidirectional path}($V, i, j$)
      \STATE Assume multidirpath = $[x_1,\dots,x_P]$
      \STATE pathleft $\leftarrow$ reverse(multidirpath[1 to lowestancestoridx])
      \STATE pathright $\leftarrow$ multidirpath[lowestancestoridx to $P$]
      \STATE bagsize $\leftarrow [1,\dots,1]$, an array of size $P$
      \FOR{each node $k$ that is not on multidirpath}
          \STATE bagidxleft $\leftarrow$ \textbf{Find bag}(pathleft, $k$)
          \STATE bagidxright $\leftarrow$ \textbf{Find bag}(pathright, $k$)
          \IF {bagidxleft = lowestancestoridx}
              \STATE bagidx $\leftarrow$ bagidxright + length(pathleft) $-$ 1
          \ELSE
              \STATE bagidx $\leftarrow$ bagidxleft
          \ENDIF
          \STATE bagsize[bagidx] $\leftarrow$ bagsize[bagidx] + 1
      \ENDFOR
      \STATE leftsize $\leftarrow$ 0
      \STATE evenseparator $\leftarrow$ None
      \FOR{$r = 1,\dots,P-1$}
          \STATE leftsize $\leftarrow$ leftsize + bagsize[r]
          \IF{leftsize $ \in [n/d, (d-1)n/d]$ }
              \STATE evenseparator $\leftarrow$ $(x_r, x_{r+1})$
              \BREAK
          \ENDIF
      \ENDFOR
      \IF {evenseparator $\neq$ None}
          \BREAK
      \ENDIF
  \ENDWHILE
  \STATE ($V_1$, $V_2$) $\leftarrow$ \textbf{Split tree}($V$, evenseparator)
  \STATE $E_1$ $\leftarrow$ \textbf{Reconstruct tree}($V_1$)
  \STATE $E_2$ $\leftarrow$ \textbf{Reconstruct tree}($V_2$)
  \STATE $E \leftarrow E_1 \cup E_2 \cup \{$ evenseparator $\}$
\ENDIF
\STATE \textbf{Output:} edge set $E$
\end{algorithmic}
\end{algorithm}

Next, we present Algorithm~\ref{alg:findmultidirpath} for recovering the sequence of nodes in the multidirectional path between any two nodes.

\begin{algorithm}[H]
\caption{Reconstruct multidirectional path}
\label{alg:findmultidirpath}
\begin{algorithmic}[1]
\STATE \textbf{Input:} vertex set $V$ of the directed rooted tree $T$, two nodes $i, j \in V$
\STATE multidirpath = [], lowestancestoridx = None
\IF{$Q(i, j) = 1$}
    \FOR{each node $k \in V - \{i,j\}$ where $Q(i, k) = 1$ and $Q(k, j) = 1$}
        \STATE multidirpath $\leftarrow$ append(multidirpath, $k$)
    \ENDFOR
    \STATE Sort multidirpath with the order relation given in Definition~\ref{def:orderrelation} \label{lin:findmultidirpathsort1}
    \STATE multidirpath $\leftarrow$ append($i$, multidirpath, $j$)
    \STATE lowestancestoridx $\leftarrow$ 1
\ELSIF{$Q(j, i) = 1$}
    \FOR{each node $k \in V - \{i,j\}$ where $Q(j, k) = 1$ and $Q(k, i) = 1$}
        \STATE multidirpath $\leftarrow$ append(multidirpath, $k$)
    \ENDFOR
    \STATE Sort multidirpath with the order relation given in Definition~\ref{def:orderrelation} \label{lin:findmultidirpathsort2}
    \STATE multidirpath $\leftarrow$ append($j$, multidirpath, $i$)
    \STATE lowestancestoridx $\leftarrow$ 1
\ELSE
    \STATE $m$ $\leftarrow$ \textbf{Find lowest common ancestor}($V, i, j$) \label{lin:findmultidirpathancestor}
    \STATE pathleft $\leftarrow$ [], pathright $\leftarrow$ []
    \FOR{each node $k \in V - \{m,i\}$ where $Q(m, k) = 1$ and $Q(k, i) = 1$}
        \STATE pathleft $\leftarrow$ append(pathleft, $k$)
    \ENDFOR
    \STATE Sort pathleft with the order relation given in Definition~\ref{def:orderrelation} \label{lin:findmultidirpathsort3}
    \FOR{each node $k \in V - \{m,j\}$ where $Q(m, k) = 1$ and $Q(k, j) = 1$}
        \STATE pathright $\leftarrow$ append(pathright, $k$)
    \ENDFOR
    \STATE Sort pathright with the order relation given in Definition~\ref{def:orderrelation} \label{lin:findmultidirpathsort4}
    \STATE multidirpath $\leftarrow$ append($i$, reverse(pathleft), $m$, pathright, $j$)
    \STATE lowestancestoridx $\leftarrow$ 2+length(pathleft)
\ENDIF
\STATE \textbf{Output:} multidirpath, lowestancestoridx
\end{algorithmic}
\end{algorithm}

In what follows, we present Algorithm~\ref{alg:findcommonancestor} for finding the lowest common ancestor of a multidirectional path between any two arbitrary nodes.

\begin{algorithm}[H]
\caption{Find lowest common ancestor}
\label{alg:findcommonancestor}
\begin{algorithmic}[1]
\STATE \textbf{Input:} vertex set $V$ of the directed rooted tree $T$, two nodes $i, j \in V$
\STATE pathfromroot $\leftarrow$ \textbf{Find path from root}($V$, $i$) \label{lin:findcommonancestorroot}
\FOR{each node $k$ in pathfromroot}
    \IF{$Q(k, j) = 1$}
        \STATE lowestcommonancestor $\leftarrow k$
        \BREAK
    \ENDIF
\ENDFOR
\STATE \textbf{Output:} lowestcommonancestor
\end{algorithmic}
\end{algorithm}

Next, we present Algorithm~\ref{alg:findroot} for identifying the path from the root to a given arbitrary node.

\begin{algorithm}[H]
\caption{Find path from root}
\label{alg:findroot}
\begin{algorithmic}[1]
\STATE \textbf{Input:} vertex set $V$ of the directed rooted tree $T$, node $i \in V$
\STATE pathfromroot $\leftarrow$ []
\FOR{each node $j \in V - \{i\}$}
    \IF{$Q(j, i) = 1$}
        \STATE pathfromroot $\leftarrow$ append(pathfromroot, $j$)
    \ENDIF
\ENDFOR
\STATE Sort pathfromroot with the order relation given in Definition~\ref{def:orderrelation} \label{lin:findrootsort}
\STATE \textbf{Output:} pathfromroot
\end{algorithmic}
\end{algorithm}

Next, we present Algorithm~\ref{alg:findbag} for finding out the bag of a node with respect to an arbitrary directed path.

\begin{algorithm}[H]
\caption{Find bag}
\label{alg:findbag}
\begin{algorithmic}[1]
\STATE \textbf{Input:} path $x_1, x_2,\dots, x_k$, node $i \in V$ where $V$ is the vertex set of the directed rooted tree $T$
\STATE $l \leftarrow 1$, $r \leftarrow k$
\WHILE{$l < r$}
    \STATE $m \leftarrow (l+r)/2$
    \IF{$Q(x_m, i) = 1$}
        \STATE $l \leftarrow m$
    \ELSE
        \STATE $r \leftarrow m$
    \ENDIF
\ENDWHILE
\STATE \textbf{Output:} bag index $l \in \{1, \dots, k\}$
\end{algorithmic}
\end{algorithm}

In what follows, we present Algorithm~\ref{alg:splittree} for splitting a directed rooted tree into two subtrees, by cutting the original tree through any arbitrary edge.

\begin{algorithm}[H]
\caption{Split tree}
\label{alg:splittree}
\begin{algorithmic}[1]
\STATE \textbf{Input:} vertex set $V$ of the directed rooted tree $T=(V,E)$, edge $(i,j) \in E$
\STATE $V_1 \leftarrow \emptyset, V_2 \leftarrow \{j\}$
\FOR{each node $k \in V - \{j\}$} \label{lin:splittreefor}
    \IF{$Q(j, k) = 1$}
        \STATE $V_2 \leftarrow V_2 \cup \{k\}$
    \ELSE
        \STATE $V_1 \leftarrow V_1 \cup \{k\}$
    \ENDIF
\ENDFOR
\STATE \textbf{Output:} partitions $V_1$, $V_2$ of $V$
\end{algorithmic}
\end{algorithm}

Next, we present Algorithm~\ref{alg:reconstructnoisytree} for reconstructing a directed rooted tree by using \emph{noisy} path queries.
We will use $\Qn\si{k}(i,j)$ to denote the $k$-th call to the query $\Qn_T(i,j)$, since for our problem, the directed rooted tree $T$ is fixed (but unknown).

\begin{algorithm}[H]
\caption{Reconstruct tree from noisy queries}
\label{alg:reconstructnoisytree}
\begin{algorithmic}[1]
\STATE \textbf{Input:} vertex set $V$ of the directed rooted tree $T$, number of queries $m$ per node pair
\STATE Define the \emph{path query} $Q'$ as follows.
Let $Q'(i,j) = 1$ if $\sum_{k=1}^m\Qn\si{k}(i,j) > m/2$, and $Q'(i,j) = 0$ otherwise.
\STATE $E$ $\leftarrow$ \textbf{Reconstruct tree}($V$) \label{lin:reconstructnoisytreecall}
\STATE \textbf{Output:} edge set $E$
\end{algorithmic}
\end{algorithm}

Finally, we present Algorithm~\ref{alg:reconstructweightedtree} for reconstructing a weighted directed rooted tree by using \emph{additive} path queries.
We will use $\Qa(i,j)$ to denote $\Qa_T(i,j)$ since for our problem, the weighted directed rooted tree $T$ is fixed (but unknown).

\begin{algorithm}[H]
\caption{Reconstruct weighted tree from additive queries}
\label{alg:reconstructweightedtree}
\begin{algorithmic}[1]
\STATE \textbf{Input:} vertex set $V$ of the weighted directed rooted tree $T$
\STATE Define the \emph{path query} $Q$ as follows.
Let $Q(i,j) = 1$ if $\Qa(i,j) > 0$, and $Q(i,j) = 0$ otherwise.
\STATE $E$ $\leftarrow$ \textbf{Reconstruct tree}($V$) \label{lin:reconstructweightedtreecall}
\STATE $W \leftarrow 0$
\FOR{ each edge $(i,j) \in E$ }
\STATE $w_{i,j}$ = $\Qa(i,j)$
\ENDFOR
\STATE \textbf{Output:} edge set $E$, edge weights $W$
\end{algorithmic}
\end{algorithm}

\section{Detailed Proofs} \label{sec:detailedproofs}

In this section, we state the proofs of all the theorems and lemmas in our manuscript.

\subsection{Proof of Lemma~\ref{lem:multidirpathfortrees}}

\begin{proof}
Existence and uniqueness follows straightforwardly from the fact that the directed rooted tree $T$ is \emph{weakly} connected (i.e., the undirected skeleton graph of $T$ is connected.)
For the second claim, note that each node in $T=(V,E)$ has at most one parent.
Given any three different nodes $p,q,r \in V$, the following condition holds $\neg ( (p,r) \in E \, \wedge \, (q,r) \in E )$.
Thus, we can only construct the three cases provided in the statement, otherwise we would violate the latter condition.
\qedhere
\end{proof}

\subsection{Proof of Theorem~\ref{thm:complexity}}

\begin{proof}
Let a ``round'' be a repetition of the ``while'' loop at Line~\ref{lin:round} of Algorithm~\ref{alg:reconstructtree}.
We show that there are $\O(d)$ rounds in expectation, and that each round takes $\O(n \log n)$ time.
We then finish the proof by the application of the master theorem.

First, we analyze the expected number of rounds for finding an even separator \emph{in the multidirectional path} between two nodes chosen independently and uniformly at random.
Recall that by Lemma~\ref{lem:multidirpathfortrees} for directed rooted trees, a multidirectional path between any two arbitrary nodes always exists and is unique.
Thus, it remains to analyze the probability that two randomly chosen nodes lie on a different subtree defined by the even separator.

Recall that Corollary 2.3 in~\cite{Chung90} shows if the tree $T$ has node degree at most $d$, then an even separator exists.
More specifically, from Definition~\ref{def:evenseparator}, we know the even separator splits the tree into two subtrees, where each of the subtrees have a number of nodes between $n/d$ and $(d-1)n/d$.
Next, we reason about the two randomly selected nodes on each round.
Let $q$ be the proportion of nodes on the first subtree, and $1-q$ be the proportion of nodes on the second subtree.
We know that $q \in [1/d, (d-1)/d]$ and similarly $1-q \in [1/d, (d-1)/d]$.
Since both nodes are selected independently and uniformly at random from the set of $n$ nodes, then $q^2$ is the probability that both nodes fall in the first subtree.
Similarly, $(1-q)^2$ is the probability that both nodes fall in the second subtree.
The probability $p$ that the two nodes lie on a different subtree is
\begin{align*}
p & \geq \min_{q \in [1/d, (d-1)/d]}{\left( 1-q^2-(1-q)^2 \right)} \\
 & = 1 - (1/d)^2 - ((d-1)/d)^2 \\
 & = 2 (d-1) / d^2 \\
 & \in \Omega(1/d) \; .
\end{align*}
Therefore, the expected number of rounds $r$ until we successfully find two nodes lying on a different subtree is $E[r] = \sum_{r=1}^\infty{r(1-p)^{r-1}p} = 1/p$ which is $\O(d)$.
We now show that each round takes $\O(n \log n)$ time.
First, we derive a sequence of conclusions regarding the subroutines:
\begin{itemize}
\item Algorithm~\ref{alg:findroot} takes $\O(n\log n)$ time, since the most time-consuming step is sorting in Line~\ref{lin:findrootsort}.
\item Algorithm~\ref{alg:findcommonancestor} takes $\O(n\log n)$ time, since the most time-consuming step is the call to Algorithm~\ref{alg:findroot} in Line~\ref{lin:findcommonancestorroot}, which takes $\O(n\log n)$ time.
\item Algorithm~\ref{alg:findmultidirpath} takes $\O(n\log n)$ time, since the most time-consuming steps are sorting in Lines~\ref{lin:findmultidirpathsort1}, \ref{lin:findmultidirpathsort2}, \ref{lin:findmultidirpathsort3} and \ref{lin:findmultidirpathsort4}, and the call to Algorithm~\ref{alg:findcommonancestor} in Line~\ref{lin:findmultidirpathancestor}, which takes $\O(n\log n)$ time.
\item Algorithm~\ref{alg:findbag} takes $\O(\log n)$ time, since it performs binary search.
\item Algorithm~\ref{alg:splittree} takes $\O(n)$ time, since the ``for'' loop in Line~\ref{lin:splittreefor} iterates for at most $n$ times.
\end{itemize}
Recall that each round calls Algorithms~\ref{alg:findmultidirpath}, \ref{alg:findbag} and \ref{alg:splittree}.
It can be observed then, that the most time-consuming step on each round is the call to Algorithm~\ref{alg:findmultidirpath}, which takes $\O(n \log n)$ time.

Thus, so far we know that the time complexity of the ``while'' loop at Line~\ref{lin:round} is $\O(dn \log n)$.
To finalize the proof, note that Algorithm~\ref{alg:reconstructtree} exits the ``while'' loop at Line~\ref{lin:round} when it finds an even separator.
Recall from Definition~\ref{def:evenseparator} that the even separator splits the tree into two subtrees, where each of the subtrees have a number of nodes between $n/d$ and $(d-1)n/d$.
The total running time for Algorithm~\ref{alg:reconstructtree} is given by the recursive formula
\begin{align*}
C(n) = C(n/d) + C((d-1)n/d) + \O(dn \log n) \; .
\end{align*}
For clarity, we rewrite $C(n)$ in terms of the master theorem in~\cite{Akra98}.
That is, $C(n) = \alpha_1 C(\beta_1 n) + \alpha_2 C(\beta_2 n) + \gamma(n)$, for $\alpha_1 = \alpha_2 = 1$, $\beta_1 = 1/d$, $\beta_2 = (d-1)/d$ and $\gamma(n) \in \O(dn \log n)$.
By invoking the master theorem in~\cite{Akra98}, we have that $C(n) \in \O(n^s \int_1^n{ \gamma(z) / z^{s+1} dz })$, where $s$ is the value for which $\alpha_1 \beta_1^s + \alpha_2 \beta_2^s = 1$.
In our case $s=1$ and thus $C(n) \in \O(dn \log^2 n)$.

For the remainder of the proof, we will use $C$ to denote $C(n)$ since $n$ is a constant.
Note that $C$ is a non-negative random variable with expectation $\E[C] \in \O(dn \log^2 n)$.
By Markov's inequality, we have that $\P[C > a] \leq \E[C]/a$.
By letting $a = \E[C]/\delta$ we have $\P[C > \E[C]/\delta] \leq \delta$.
Therefore $\P[C \leq \E[C]/\delta] \geq 1-\delta$, and we prove our claim that $\P[C \in \O(\frac{1}{\delta} \, dn \log^2 n)] \geq 1-\delta$.
\qedhere
\end{proof}

\subsection{Proof of Theorem~\ref{thm:lowerboundrandomized}}

\begin{proof}
Here we provide an information-theoretic lower bound for any \emph{randomized} algorithm, based on Fano's inequality~\cite{Cover06}.

Let $\T_{n,d}$ be the set of directed rooted trees of $n$ nodes and node degree at most $d$.
Next, we show that $\log|\T_{n,d}| \in \Theta(n\log n)$.
Interestingly, the latter (tight) bound does not depend on $d$.
For a lower bound, note that the number of directed rooted trees with at most one children is equal to the number of permutations of $n$ nodes, which is $n!$.
We have $\log|\T_{n,d}| \geq \log n! \geq n(\log n - 1) \geq \frac{1}{4} n\log n$ for $n \geq 4$ and therefore $\log|\T_{n,d}| \in \Omega(n\log n)$.
For an upper bound, note that the number of directed rooted trees with no constraint in the number of children is equal to the number of directed spanning trees, which is $n^{n-1}$ by Cayley's formula.
We have $\log|\T_{n,d}| \leq \log n^{n-1} \leq n\log n$ and therefore $\log|\T_{n,d}| \in \O(n\log n)$.
From the above, we conclude that $\log|\T_{n,d}| \in \Theta(n\log n)$.

Assume that nature picks a directed rooted tree $T^* \in \T_{n,d}$ uniformly at random.
Any mechanism for finding out the correct tree is allowed to make $C$ possibly-dependent queries (for $C$ node pairs) for which the binary responses are $Q_1,\dots,Q_C$.
Let $T \in \T_{n,d}$ be the directed rooted tree that is guessed on the basis of the $C$ query responses.
The above defines a Markov chain $T^* \rightarrow (Q_1,\dots,Q_C) \rightarrow T$.

By properties of the mutual information, and since the joint responses $(Q_1,\dots,Q_C)$ can take up to $2^C$ possible values, we have:
\begin{align*}
\I(T^*;Q_1,\dots,Q_C) & \leq \H(Q_1,\dots,Q_C) \\
 & \leq C\log 2 \; .
\end{align*}
By the Fano's inequality~\cite{Cover06} on the Markov chain $T^* \rightarrow (Q_1,\dots,Q_C) \rightarrow T$ we have
\begin{align*}
\P[T \neq T^*] & \geq 1 - \frac{ \I(T^*;Q_1,\dots,Q_C) + \log{2} }{ \log{|\T_{n,d}|} } \\
 & \geq 1 - \frac{ C\log 2 + \log{2} }{ n(\log n - 1) } \\
 & \equiv \delta \; .
\end{align*}
By solving for $C$, we have that if $C \leq (1-\delta) \, n(\log n - 1)/\log 2 - 1$ then $\P[T \neq T^*] \geq \delta$ and we prove our claim.
\qedhere
\end{proof}

\subsection{Proof of Theorem~\ref{thm:lowerbounddeterministic}}

\begin{proof}
The proof relies on constructing a family of ``parallel-chain'' trees.
We assume that the algorithm knows that the directed rooted tree to be reconstructed is ``parallel-chain''.
We also assume that the algorithm has additional side information (described later for clarity.)

Assume that $n-1$ is divisible by $d$, and let $k=\frac{n-1}{d}$.
Assume the node set $V$ is partitioned into $k+1$ fixed sets $V_0,V_1,\dots,V_k$, such that $|V_0|=1$ and $|V_l|=d$ for $l=1,\dots,k$.
We then create a tree $T=(V,E)$ with $kd$ edges such that: the single node in $V_0$ has $d$ children (each node in $V_1$), $n-d-1$ nodes have one child (each node in $V_l$ has one children in $V_{l+1}$ for  $l=1,\dots,k-1$), and the remaining $d$ nodes (in $V_k$) are leaves.
The above creates a ``parallel-chain'' directed rooted tree as shown in Figure~\ref{fig:parallelchaintree}.

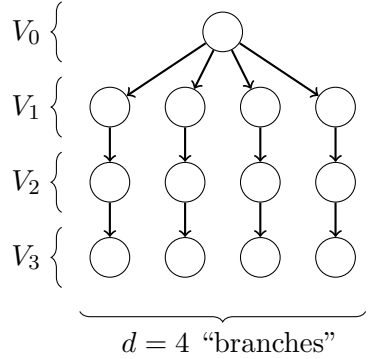
\begin{figure}[H]
\begin{center}
\begin{tikzpicture}[scale=1.0]
\tikzstyle{vertex}=[circle, fill=white, draw, inner sep=0pt, minimum size=15pt]
\node[vertex](x0) at (2.5,4) {};
\node[vertex](x1l1) at (1,3) {};
\node[vertex](x2l1) at (2,3) {};
\node[vertex](x3l1) at (3,3) {};
\node[vertex](x4l1) at (4,3) {};
\node[vertex](x1l2) at (1,2) {};
\node[vertex](x2l2) at (2,2) {};
\node[vertex](x3l2) at (3,2) {};
\node[vertex](x4l2) at (4,2) {};
\node[vertex](x1l3) at (1,1) {};
\node[vertex](x2l3) at (2,1) {};
\node[vertex](x3l3) at (3,1) {};
\node[vertex](x4l3) at (4,1) {};
\tikzset{EdgeStyle/.style={->}}
\Edge(x0)(x1l1)
\Edge(x0)(x2l1)
\Edge(x0)(x3l1)
\Edge(x0)(x4l1)
\Edge(x1l1)(x1l2)
\Edge(x2l1)(x2l2)
\Edge(x3l1)(x3l2)
\Edge(x4l1)(x4l2)
\Edge(x1l2)(x1l3)
\Edge(x2l2)(x2l3)
\Edge(x3l2)(x3l3)
\Edge(x4l2)(x4l3)
\draw[decorate,decoration={brace,amplitude=4pt},xshift=-4pt,yshift=0pt]
(0.5,3.6) -- (0.5,4.4) node[midway,left,xshift=-4pt] 
{$V_0$};
\draw[decorate,decoration={brace,amplitude=4pt},xshift=-4pt,yshift=0pt]
(0.5,2.6) -- (0.5,3.4) node[midway,left,xshift=-4pt] 
{$V_1$};
\draw[decorate,decoration={brace,amplitude=4pt},xshift=-4pt,yshift=0pt]
(0.5,1.6) -- (0.5,2.4) node[midway,left,xshift=-4pt] 
{$V_2$};
\draw[decorate,decoration={brace,amplitude=4pt},xshift=-4pt,yshift=0pt]
(0.5,0.6) -- (0.5,1.4) node[midway,left,xshift=-4pt] 
{$V_3$};
\draw[decorate,decoration={brace,amplitude=4pt},xshift=0pt,yshift=8pt]
(4.4,0) -- (0.6,0) node[below,align=center,xshift=2cm,yshift=-4pt] 
{$d=4$ ``branches''};
\end{tikzpicture}
\end{center}
\vspace{-0.25in}
\caption{A ``parallel-chain'' directed rooted tree.}
\label{fig:parallelchaintree}
\end{figure}

Moreover, we assume that the algorithm also knows the node sets $V_0,V_1,\dots,V_k$.
What remains to be recovered is the ``branch'' (from $1$ to $d$) to which the nodes in $V_1,\dots,V_k$ belong to.
In order to do this, the algorithm has to ask a query $Q(i,j)$ for each $i \in V_l$, $j \in V_{l+1}$ and  $l=1,\dots,k-1$.
The total number of queries is $(k-1)d^2 = (\frac{n-1}{d}-1)d^2 \geq \frac{dn}{2}$ for $n \geq 2(d+1)$.
Thus, a deterministic algorithm requires at least $\frac{dn}{2}$ queries.
\qedhere
\end{proof}

\subsection{Proof of Theorem~\ref{thm:complexitynoisy}}

\begin{proof}
Our first goal is to find the condition for which the path query $Q$ is equal to its noisy approximation $Q'$.
We start by making some observations for a single fixed node pair $(i,j)$ and later extends our observations for several node pairs.

Define the random variable $R(i,j) \equiv \frac{1}{m} \sum_{k=1}^m\Qn\si{k}(i,j)$.
Recall that we defined the \emph{path query} $Q'$ as follows.
Let $Q'(i,j) = 1$ if $R(i,j) > 1/2$, and $Q'(i,j) = 0$ otherwise.

Assume that $Q(i,j) = 1$.
By Definition~\ref{def:noisyquery}, we have that $\Qn\si{k}(i, j) = 1$ with probability $1-\varepsilon$, and $\Qn\si{k}(i, j) = 0$ with probability $\varepsilon$.
Clearly, $\E[\Qn\si{k}(i, j)] = 1-\varepsilon$ and therefore $\E[R(i,j)] = 1-\varepsilon$.
By using Hoeffding's inequality, we have
\begin{align*}
\P[Q(i,j) \neq Q'(i,j)] & = \P[R(i,j) < 1/2] \\
 & = \P[R(i,j) - \E[R(i,j)] < 1/2 - \E[R(i,j)]] \\
 & = \P[R(i,j) - \E[R(i,j)] < \varepsilon - 1/2] \\
 & \leq \exp{-2m(1/2-\varepsilon)^2} \; .
\end{align*}

Now assume that $Q(i,j) = 0$.
By Definition~\ref{def:noisyquery}, we have that $\Qn\si{k}(i, j) = 0$ with probability $1-\varepsilon$, and $\Qn\si{k}(i, j) = 1$ with probability $\varepsilon$.
Clearly, $\E[\Qn\si{k}(i, j)] = \varepsilon$ and therefore $\E[R(i,j)] = \varepsilon$.
By using Hoeffding's inequality, we have
\begin{align*}
\P[Q(i,j) \neq Q'(i,j)] & = \P[R(i,j) > 1/2] \\
 & = \P[R(i,j) - \E[R(i,j)] > 1/2 - \E[R(i,j)]] \\
 & = \P[R(i,j) - \E[R(i,j)] > 1/2 - \varepsilon] \\
 & \leq \exp{-2m(1/2-\varepsilon)^2} \; .
\end{align*}

Assume that Algorithm~\ref{alg:reconstructtree} in Line~\ref{lin:reconstructnoisytreecall} of Algorithm~\ref{alg:reconstructnoisytree} makes $C$ path queries.
That is, assume that queries were made for $C$ node pairs $(i_1,j_1),(i_2,j_2),\dots,(i_C,j_C)$.
By the union bound and the previous observations, we have:
\begin{align*}
\P[(\exists k = 1,\dots,C) \, Q(i_k,j_k) \neq Q'(i_k,j_k)] & \leq C \exp{-2m(1/2-\varepsilon)^2} \\
 & \equiv \delta/2 \; .
\end{align*}
By solving for $m$, we have that if $m \geq \frac{1}{2(1/2-\varepsilon)^2} \, (\log C + \log\frac{2}{\delta})$ then
\begin{align*}
\P[(\forall k = 1,\dots,C) \, Q(i_k,j_k) = Q'(i_k,j_k)] \geq 1-\delta/2 \; .
\end{align*}
That is, with probability at least $1-\delta/2$, the path query $Q$ is equal to its noisy approximation $Q'$ and thus Algorithm~\ref{alg:reconstructnoisytree} reconstructs the tree $T$ correctly, provided that $m$ is large enough.

By Theorem~\ref{thm:complexity}, with probability at least $1-\delta/2$, we have that $C \in \O(\frac{2}{\delta} \, dn \log^2 n)$.
From the above, we have that $m \in \Theta(\frac{1}{(1/2-\varepsilon)^2} \, (\log d + \log n + \log\frac{1}{\delta}))$ fulfills $m \geq \frac{1}{2(1/2-\varepsilon)^2} \, (\log C + \log\frac{2}{\delta})$.
Finally, the total number of queries is given by $C m \in \O(\frac{1}{\delta} \, \frac{1}{(1/2-\varepsilon)^2} \, dn \log^2 n (\log d + \log n + \log\frac{1}{\delta}))$ and we prove our claim.
\qedhere
\end{proof}

\subsection{Proof of Theorem~\ref{thm:complexityweighted}}

\begin{proof}
Straightforwardly, by the fact that the most time-consuming step in Algorithm~\ref{alg:reconstructweightedtree} is the call to Algorithm~\ref{alg:reconstructtree} in Line~\ref{lin:reconstructweightedtreecall}, and by the result in Theorem~\ref{thm:complexity}.
\qedhere
\end{proof}

\subsection{Proof of Theorem~\ref{thm:lowerbounddags}}

\begin{proof}
Let $\G_n$ be the set of directed acyclic graphs of $n$ nodes, without transitive edges.
Next, we show that $\log|\G_n| \in \Theta(n^2)$.
For a lower bound, we focus on a particular class of ``two-layered'' graphs.
First, the node set $V=\{1,\dots,n\}$ is partitioned into two sets $V_1 = \{1,\dots,\floor{n/2}\}$ and $V_2 = \{\floor{n/2}+1,\dots,n\}$.
We then allow only for edges from nodes in $V_1$ to nodes in $V_2$.
That is, each node in $V_2$ can have as parents any subset of the nodes in $V_1$, thus, there are $2^{|V_1|}$ choices of edge sets for each of the $|V_2|$ nodes in $V_2$.
We have $\log|\G_n| \geq \log 2^{|V_1||V_2|} = \log 2^{\floor{n/2}(n-\floor{n/2})} \geq \log 2^{n(n-1)/4} \geq \frac{\log 2}{5} n^2$ for $n \geq 5$ and therefore $\log|\G_n| \in \Omega(n^2)$.
For an upper bound, note that the number of directed acyclic graphs (without transitive edges) is less than the number of directed graphs of $n$ nodes, which is $3^{\binom{n}{2}}$.
This follows from the fact that for any two nodes $i$ and $j$, we have three cases in a directed graph $G=(V,E)$.
Either $(i,j) \in E$, or $(j,i) \in E$ or $\{(i,j),(j,i)\} \not\subset E$.
Now, we have $\log|\G_n| \leq \log 3^{\binom{n}{2}} \leq \frac{\log 3}{2} n^2$ and therefore $\log|\G_n| \in \O(n^2)$.
From the above, we conclude that $\log|\G_n| \in \Theta(n^2)$.

Since each query only reveals a single bit of information, the lower bound of $\Theta(n^2)$ for any \emph{deterministic} algorithm follows.

For the lower bound for any \emph{randomized} algorithm, we proceed as in Theorem~\ref{thm:lowerboundrandomized} by using Fano's inequality~\cite{Cover06}.
\qedhere
\end{proof}

\subsection{Proof of Theorem~\ref{thm:lowerboundedge}}

\begin{proof}
The proof relies on constructing a family of graphs with a single edge.
We assume that the algorithm knows that the directed acyclic graph to be reconstructed has a single edge.

Assume two fixed nodes $i,j \in V$.
The directed graph to be reconstructed is $G=(V,E)$ where $E=\{(i,j)\}$.
Note that $Q(i,j)=1$.
Furthermore $Q(k,l)=0$ for every node pair $(k,l) \neq (i,j)$.
That is only one query returns 1, while $n^2-1$ queries return 0.
Thus, a deterministic algorithm does not obtain any information from the $n^2-1$ queries in order to guess the edge $(i,j)$, and therefore it requires at least $n^2$ queries in the worst case.
\qedhere
\end{proof}

\subsection{Proof of Theorem~\ref{thm:lowerboundsparse}}

\begin{proof}
The proof relies on constructing a family of ``v-structured two-layered'' graphs.
We assume that the algorithm only knows that the directed acyclic graph to be reconstructed has $n-1$ edges (i.e., the algorithm does not know that the graph is ``v-structured two-layered''.)

Assume the node set $V$ is partitioned into two fixed sets $V_1$ and $V_2$.
For simplicity assume that there is an odd number of nodes, and that $|V_2|=|V_1|+1$.
Thus, $|V_1| = \floor{n/2}$ and $|V_2| = \floor{n/2} + 1$.
We then create the graph $G=(V,E)$ with $n-1$ edges such that each node in $V_1$ is the source of 2 edges, and each node in $V_2$ is the target of at most 2 edges.
The above creates a ``v-structured two-layered'' graph as shown in Figure~\ref{fig:lowerboundsparse}.

\begin{figure}[H]
\begin{center}
\begin{tikzpicture}[scale=1.0]
\tikzstyle{vertex}=[circle, fill=white, draw, inner sep=0pt, minimum size=15pt]
\node[vertex](x0) at (1,1) {};
\node[vertex](x1) at (2,1) {};
\node[vertex](x2) at (3,1) {};
\node[vertex](x3) at (4,1) {};
\node[vertex](x4) at (0.5,0) {};
\node[vertex](x5) at (1.5,0) {};
\node[vertex](x6) at (2.5,0) {};
\node[vertex](x7) at (3.5,0) {};
\node[vertex](x8) at (4.5,0) {};
\tikzset{EdgeStyle/.style={->}}
\Edge(x0)(x4)
\Edge(x0)(x5)
\Edge(x1)(x5)
\Edge(x1)(x6)
\Edge(x2)(x6)
\Edge(x2)(x7)
\Edge(x3)(x7)
\Edge(x3)(x8)
\draw[decorate,decoration={brace,amplitude=4pt},xshift=-4pt,yshift=0pt]
(0,0.6) -- (0,1.4) node[midway,left,xshift=-4pt] 
{$V_1$};
\draw[decorate,decoration={brace,amplitude=4pt},xshift=-4pt,yshift=0pt]
(0,-0.4) -- (0,0.4) node[midway,left,xshift=-4pt] 
{$V_2$};
\end{tikzpicture}
\end{center}
\vspace{-0.25in}
\caption{A ``v-structured two-layered'' directed acyclic graph.}
\label{fig:lowerboundsparse}
\end{figure}
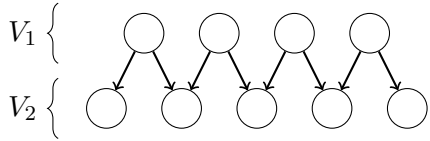

Note that $Q(i,j)=1$ for $(i,j) \in E$, while $Q(i,j)=0$ for $(i,j) \notin E$.
That is only $n-1$ queries return 1, while $n^2-n+1$ queries return 0.
Thus, a deterministic algorithm does not obtain any information from the $n^2-n+1$ queries in order to guess the edge set $E$, and therefore it requires at least $n^2-n+2$ queries in the worst case.
(Since the algorithm knows that there are $n-1$ edges, it can stop asking queries as soon as the first bit $1$ is returned.)
\qedhere
\end{proof}

\section{Experiments} \label{sec:experiments}

In this section, we present our experimental validation.
We performed $10$ repetitions for different number of nodes, and node degrees.
For each repetition, we generate a random bounded-degree directed rooted tree, in order to test whether our Algorithm~\ref{alg:reconstructtree} can successfully recover the tree by using path queries, as well as to count the number of queries needed in practice.
We experimentally found that all trees were successfully recovered in practice.
Thus, we focused on the question regarding the number of queries needed by Algorithm~\ref{alg:reconstructtree} in practice.

\begin{figure}[H]
\begin{center}
\includegraphics[width=3in]{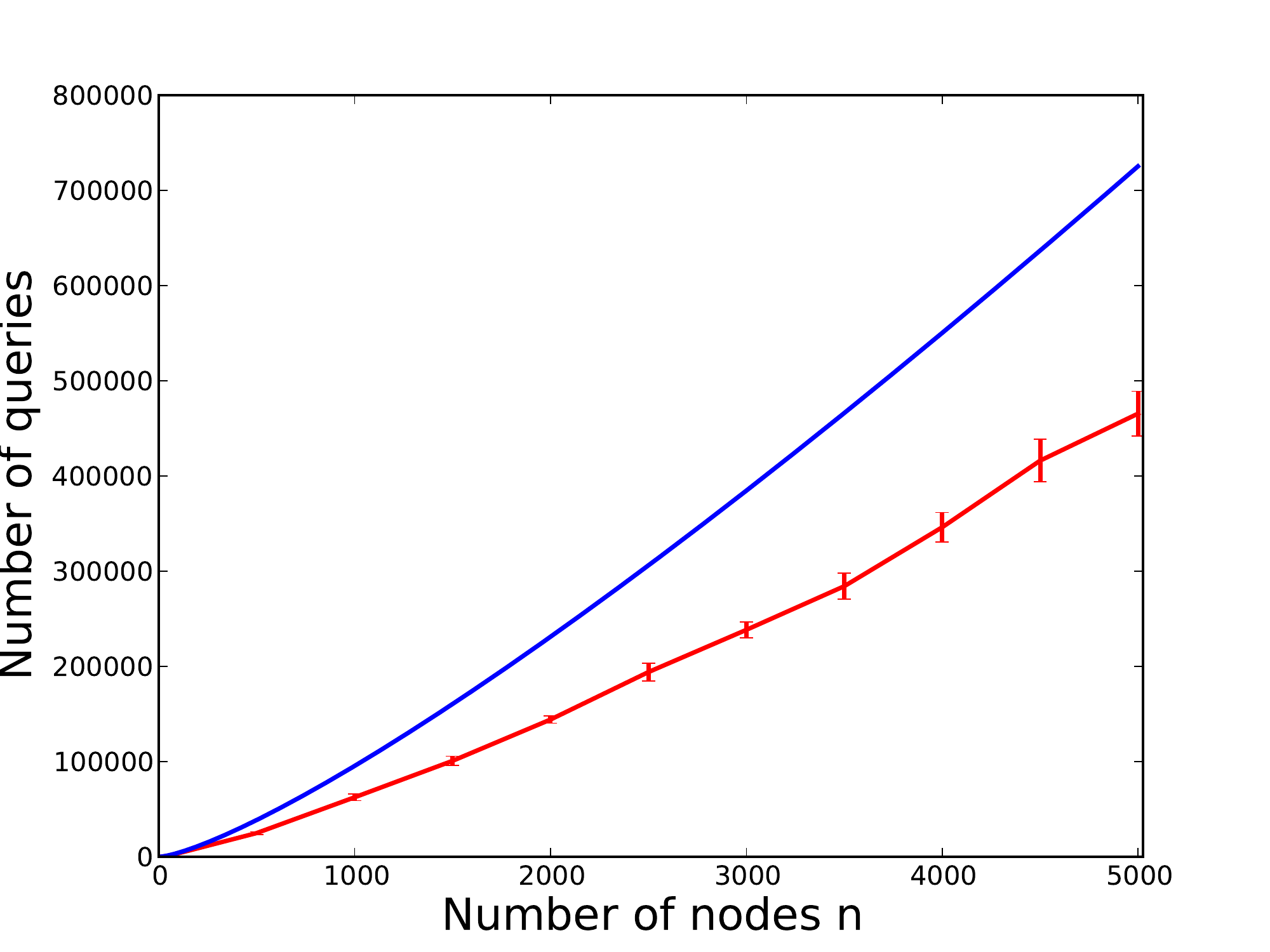}\includegraphics[width=3in]{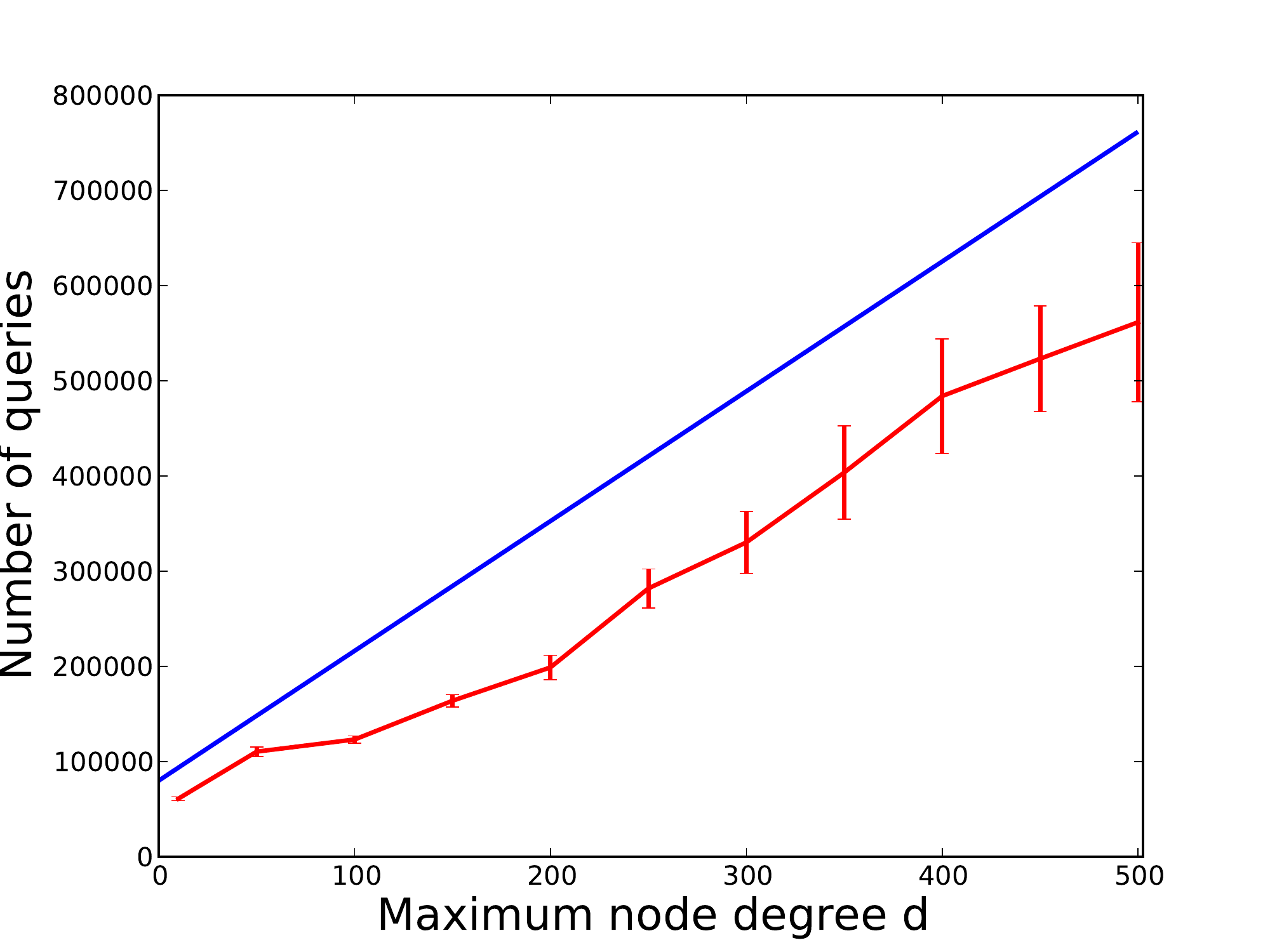} \\
\vspace{-0.2in}\makebox[1.95in]{}\makebox[1.05in]{(a)}\makebox[1.95in]{}\makebox[0.95in]{(b)}
\end{center}
\vspace{-0.25in}
\caption{(a) Number of queries for different number of nodes $n$, for a maximum node degree $d=5$.
(b) Number of queries for different node degrees $d$, for a number of nodes $n=1000$.
The upper bound for the number of queries in Theorem~\ref{thm:complexity} is shown in blue.
The actual number of queries used by Algorithm~\ref{alg:reconstructtree} is shown in red.
(Error bars were computed for $10$ repetitions, at $95\%$ significance level.)}
\label{fig:experiments}
\end{figure}

Figure~\ref{fig:experiments}(a) shows the number of queries for different number of nodes, while Figure~\ref{fig:experiments}(b) shows the number of queries for different node degrees.
From our results, it can be observed that the number of queries used by Algorithm~\ref{alg:reconstructtree} is less than what Theorem~\ref{thm:complexity} predicted.
Furthermore, for a constant degree $d$, the experimental results are much better for a large number of nodes $n$, as shown in Figure~\ref{fig:experiments}(a).

\end{document}